\documentclass[letterpaper,twocolumn,longbibliography,pre,amsmath,amssymb,float,notitlepage]{revtex4-2}
\usepackage{epsfig}
\usepackage{epstopdf}
\usepackage{float}
\usepackage[caption=false]{subfig}
\usepackage[section]{placeins}
\usepackage{graphicx}		
\usepackage{epstopdf}
\usepackage{url}
\usepackage{amsmath}
\usepackage{mathtools}
\usepackage{enumitem}
\usepackage{amssymb}
\usepackage{physics}
\usepackage{breqn}
\usepackage[hidelinks]{hyperref}
\usepackage{import}
\usepackage{tkz-euclide}
\usepackage{color}
\usepackage{soul}
\usepackage{dcolumn}
\usepackage{graphicx}
\usepackage{amsmath}
\usepackage{amssymb}
\usepackage{bm}
\usepackage{tikz}
\usepackage[section]{placeins}
\usetikzlibrary{decorations.pathmorphing,patterns}
\usepackage[capitalise]{cleveref}

\makeatletter
\let\cref@old@eq@setnumberOld\eq@setnumber
\def\eq@setnumber{%
	\cref@old@eq@setnumberOld%
	\cref@constructprefix{equation}{\cref@result}%
	\protected@xdef\cref@currentlabel{%
		[equation][\arabic{equation}][\cref@result]\p@equation\eq@number}}
		
\makeatother

\makeatletter
\let\cat@comma@active\@empty
\makeatother
\begin{document}

\title{
Breaking through millisecond coherence threshold with a Superconducting Qubit
}
\title{Millisecond coherence in a superconducting qubit}

\author{Aaron Somoroff}
\affiliation{Department of Physics, Joint Quantum Institute, and Center for Nanophysics and Advanced Materials, University of Maryland, College Park, MD 20742, USA}

\author{Quentin Ficheux}
\affiliation{Department of Physics, Joint Quantum Institute, and Center for Nanophysics and Advanced Materials, University of Maryland, College Park, MD 20742, USA}

\author{Raymond A. Mencia}
\affiliation{Department of Physics, Joint Quantum Institute, and Center for Nanophysics and Advanced Materials, University of Maryland, College Park, MD 20742, USA}

\author{Haonan Xiong}
\affiliation{Department of Physics, Joint Quantum Institute, and Center for Nanophysics and Advanced Materials, University of Maryland, College Park, MD 20742, USA}

\author{Roman Kuzmin}
\affiliation{Department of Physics, Joint Quantum Institute, and Center for Nanophysics and Advanced Materials, University of Maryland, College Park, MD 20742, USA}

\author{Vladimir E. Manucharyan}
\affiliation{Department of Physics, Joint Quantum Institute, and Center for Nanophysics and Advanced Materials, University of Maryland, College Park, MD 20742, USA}

\date{\today}

\begin{abstract}

Increasing the degree of control over physical qubits is a crucial component of quantum computing research. We report a superconducting qubit of fluxonium type with the Ramsey coherence time reaching $T_2^*= 1.48 \pm 0.13 \mathrm{~ms}$, which exceeds the state of the art value by an order of magnitude. As a result, the average single-qubit gate fidelity grew above $0.9999$, surpassing, to our knowledge, any other solid-state quantum system. Furthermore, by measuring energy relaxation of the parity-forbidden transition to second excited state, we exclude the effect of out-of-equilibrium quasiparticles on coherence in our circuit. Combined with recent demonstrations of two-qubit gates on fluxoniums, our result paves the way for the next generation of quantum processors.

\end{abstract}

\maketitle

Superconducting qubits have become a major quantum computing platform in large part because of a rapid growth of coherence time \cite{Devoret2013SuperconductingOutlook, Kjaergaard2020SuperconductingPlay}, beginning with the first demonstration of coherent oscillations in a Cooper pair box circuit in 1999 \cite{Nakamura1999CoherentBox}. Notable leaps took place with the invention of quantronium \cite{Vion2002ManipulatingCircuit} and the 3D-transmon qubit \cite{Paik2011ObservationArchitecture}, the latter leading to a widespread use of transmons and related circuits, such as X-mons \cite{Barends2013CoherentCircuits} and C-shunt flux qubits \cite{Yan2016TheReproducibility}. However, despite promising recent developments \cite{Place2020NewMilliseconds}, the coherence time of superconducting qubits measured with the Ramsey metric has been stuck at about $100~\mu\textrm{s}$ level for almost a decade \cite{Rigetti2012SuperconductingMs}. The saturation of coherence time of superconducting qubits slows down the implementation of useful intermediate-scale quantum algorithms \cite{Arute2019QuantumProcessor, Harrigan2021QuantumProcessor, Kandala2017Hardware-efficientMagnets,Hashim2020RandomizedProcessor, Song201710-QubitCircuit, Jurcevic2021DemonstrationSystem, Barends2015DigitalCircuit, Salathe2015DigitalElectrodynamics} and ultimately intensifies the hardware requirement to achieve quantum error correction \cite{Ofek2016ExtendingCircuits, Campagne-Ibarcq2020QuantumOscillator, Andersen2020RepeatedCode, Marques2021Logical-qubitCode}. Here, we report a superconducting qubit with a Ramsey coherence time above a millisecond.

\begin{figure}[t]
	\centering
	\includegraphics[width=\linewidth]{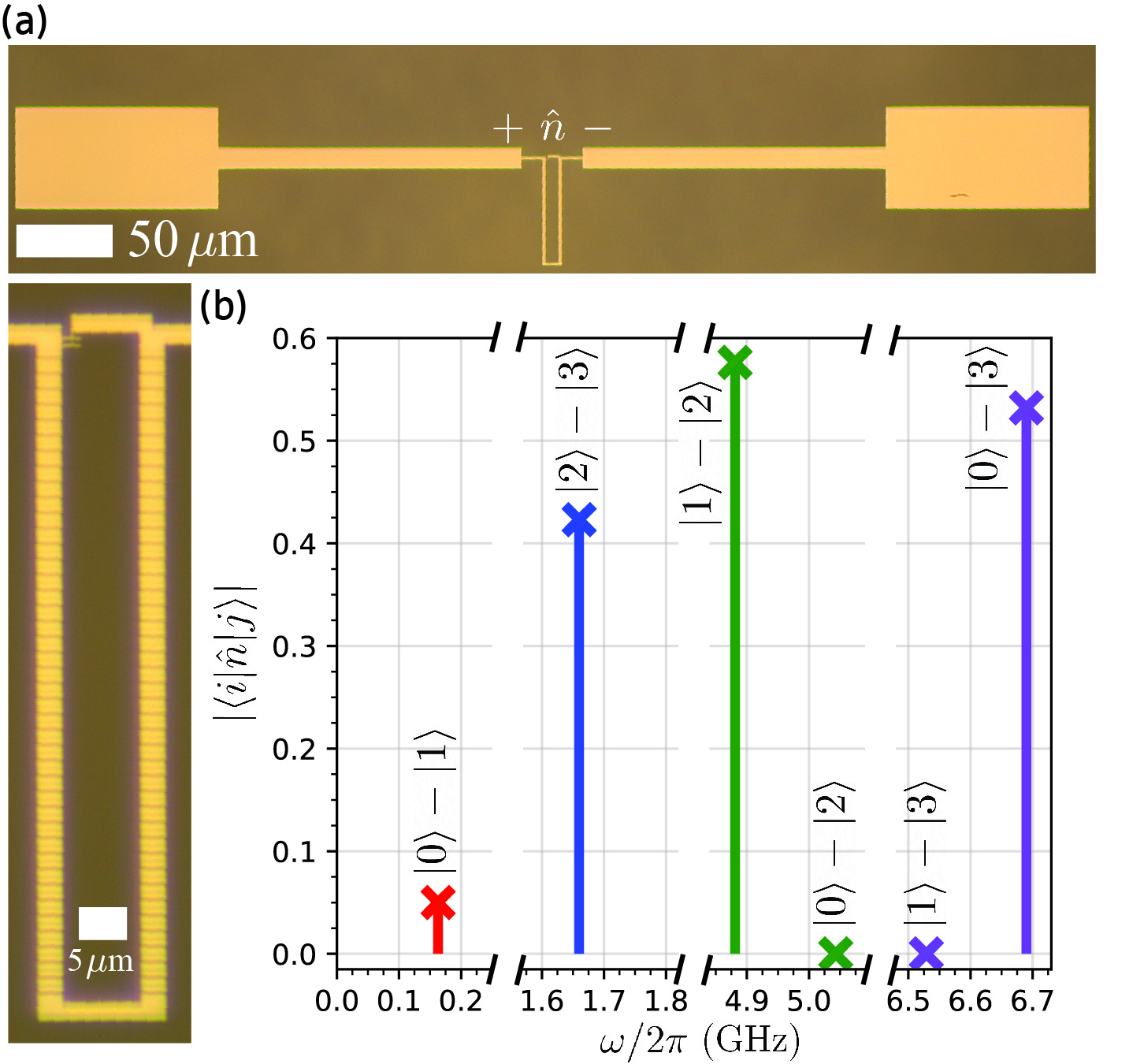}
	\caption{(a) Optical images of the device. Top panel shows the entire device. Antenna pads capacitively couple to a 3D copper cavity used for	readout. Side panel is a close up of the fluxonium circuit, showing the loop formed by the small Josephson junction (JJ) and the chain of JJ that provides the superinductance required for the fluxonium spectral regime. (b) The six lowest transitions of this device along with their charge matrix elements at half flux quantum, $\Phi_e/\Phi_0 = 0.5$.}
	\label{fig:Fig1}
\end{figure}

Our circuit, known as fluxonium~\cite{Manucharyan2009Fluxonium:Offsets}, consists of a relatively weak Josephson junction connected to an antenna-like capacitance and a compact large-value inductance (superinductance), realized with an array of about a hundred relatively strong junctions (Fig.~\ref{fig:Fig1}a). The circuit design is similar to that introduced in references \cite{Nguyen2019High-CoherenceQubit, Lin2018DemonstrationDecay} except the substrate is changed from silicon to sapphire. Fluxonium dynamics can be described using a pair of conjugate operators $\hat\varphi$ and $\hat n = -i\partial_{\varphi}$, representing, respectively, the superconducting phase-twist across the inductance and the charge displaced at the capacitor plates (in units of the Cooper pair charge). The chip is placed inside a copper cavity, with a resonance frequency of $7.54~\textrm{GHz}$ and a quality factor of $Q = 377$, in order to perform a dispersive readout of the qubit state \cite{Blais2004CavityComputation,Wallraff2004StrongElectrodynamics,Blais2020CircuitElectrodynamics}. A separate port in the cavity is used for a wireless driving of fluxonium transitions. The circuit parameters are accurately determined from spectroscopy data as a function of flux through the loop, which, along with the details of our experimental procedures, are available in the Supplementary Material.

At the half-integer flux bias, fluxoniums are practically unaffected by $1/f$ flux-noise \cite{Nguyen2019High-CoherenceQubit, Zhang2020UniversalQubit}, thanks to the large value of the inductive shunt. The spectrum of relevant transitions in the present device can be concisely summarised in Fig.~\ref{fig:Fig1}b. The qubit transition between the lowest energy states $|0\rangle$ and $|1\rangle$ has a frequency $\omega_{01}/2\pi = 163~\textrm{MHz}$. In comparison to transmons, such a qubit is better protected against energy relaxation into charge defects in the circuit material \cite{Wang2015SurfaceQubits}, because of the reduced matrix element $\langle 0|\hat n|1\rangle \ll 1$ (for transmons, $\langle 0|\hat n|1\rangle \sim 1$), and against uncontrolled leakage of quantum information into higher energy states, because of the much larger anhramonicity. The non-computational transitions $|1\rangle - |2\rangle$ and $|0\rangle - |3\rangle$ are instrumental to designing an on-demand qubit-qubit interaction \cite{Nesterov2018Microwave-activatedQubits}, as they have much larger frequency and charge matrix elements. In fact, high-fidelity controlled-Z and controlled-phase gates on a pair of fluxoniums with similar spectra to that shown in Fig.~\ref{fig:Fig1} have been recently demonstrated \cite{Ficheux2020FastFluxoniums,Xiong2021ArbitraryShifts}.

\begin{figure}[t]
	\centering
	\includegraphics[width=1.1\linewidth]{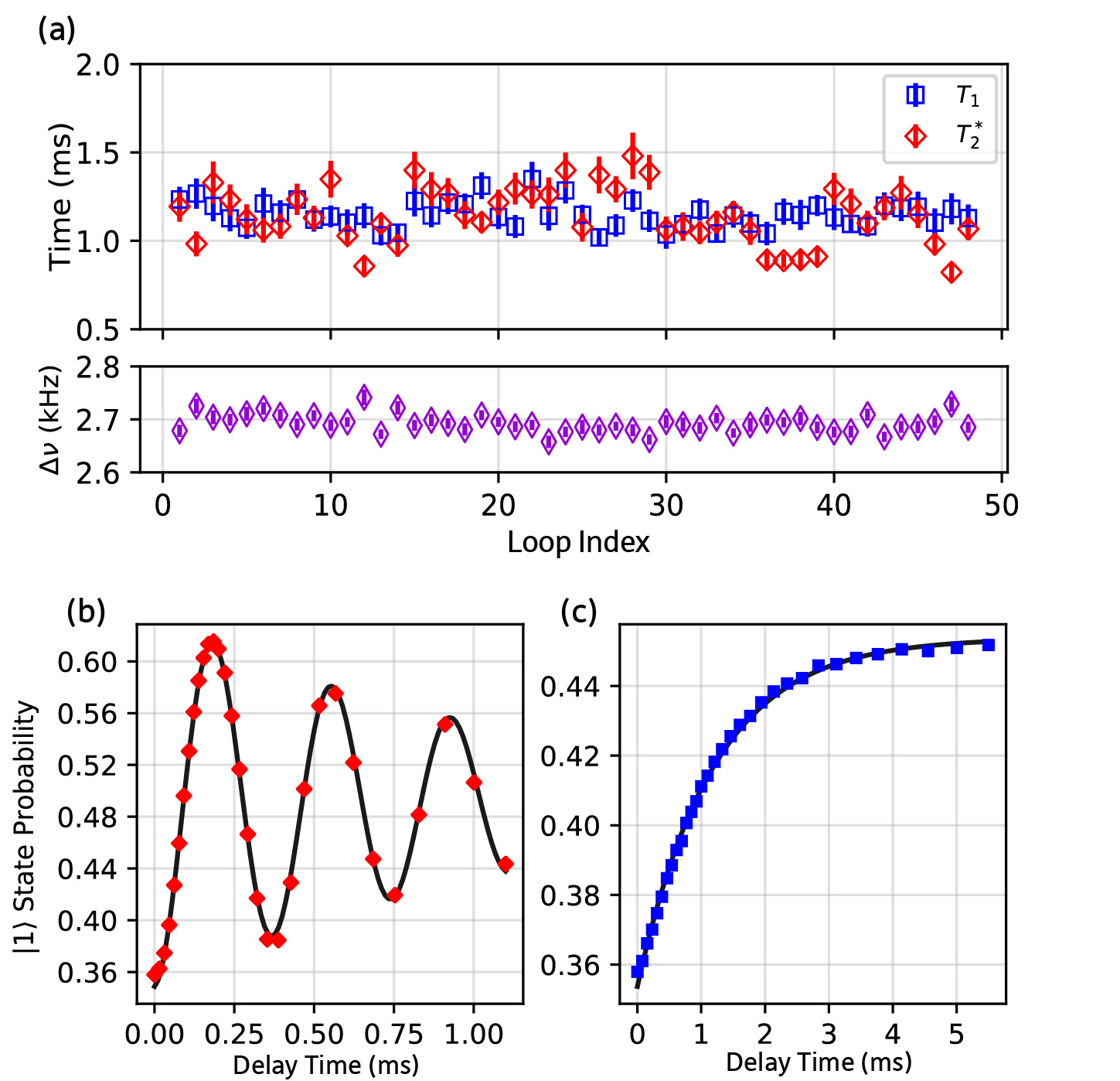}
	\caption{(a) Interleaved energy relaxation time $T_1$ and Ramsey coherence time $T_2^*$ loop over 12 hours. Below is the frequency of the Ramsey fringes. The qubit frequency is stable to within 100 Hz. (b) Ramsey fringes averaged across all 48 measurements in the loop. The solid line is the fit to a decaying sinusoid with characteristic time $\bar{T}_2^*=1.16 \pm 0.05 ~\textrm{ms}$. (c) Energy relaxation curve averaged across all measurements in the loop. The solid line is a decaying exponential fit giving $\bar{T}_1 = 1.20 \pm 0.03~\mathrm{ms}$.}
	\label{fig:Fig2}
\end{figure}

The qubit energy relaxation time $T_1$ is measured by applying a $\pi$-pulse to the $|0\rangle - |1\rangle$ transition and reading out the excited state population after a variable time delay. The acquired relaxation signal decays exponentially with a characteristic time $T_1$. Prior to the $\pi$-pulse, the qubit is initialized using the readout-induced optical pumping effect in the high photon number regime (see Supplementary Note 2C). The coherence time $T_2^*$ is obtained from the Ramsey sequence of two $\pi/2$-pulses separated by a variable time delay, without any correcting echo-pulses. This protocol produces Ramsey fringes oscillating at the drive-qubit detuning frequency $\Delta\nu$ and has an exponentially decaying envelope with a characteristic time $T_2^*$. The two pulse sequences were interleaved and repeated over a period of about 12 hours. The fit values of $T_1$, $T_2^*$ and $\Delta\nu$ are shown in Fig.~\ref{fig:Fig2}a. The highest recorded coherence time $T_2^* = 1.48 \pm 0.13~\textrm{ms}$ exceeds the state-of-the-art value for both transmons and fluxoniums by an order of magnitude. Even more strikingly, averaging the Ramsey fringes over a period of 12 hours results in only a minor reduction of the coherence time to $\bar{T}_2^*=1.16\pm 0.05~\textrm{ms}$ (Fig.~\ref{fig:Fig2}b). Likewise, averaging the energy relaxation signal shows no signs of double-exponential behavior, typical to the case of fluctuating in time decay rate (Fig.~\ref{fig:Fig2}c). In fact, the value of $T_1$ has been stable around $1~\textrm{ms}$ over a period of several months.

\begin{figure}[t]
	\centering
	\includegraphics[width=\linewidth]{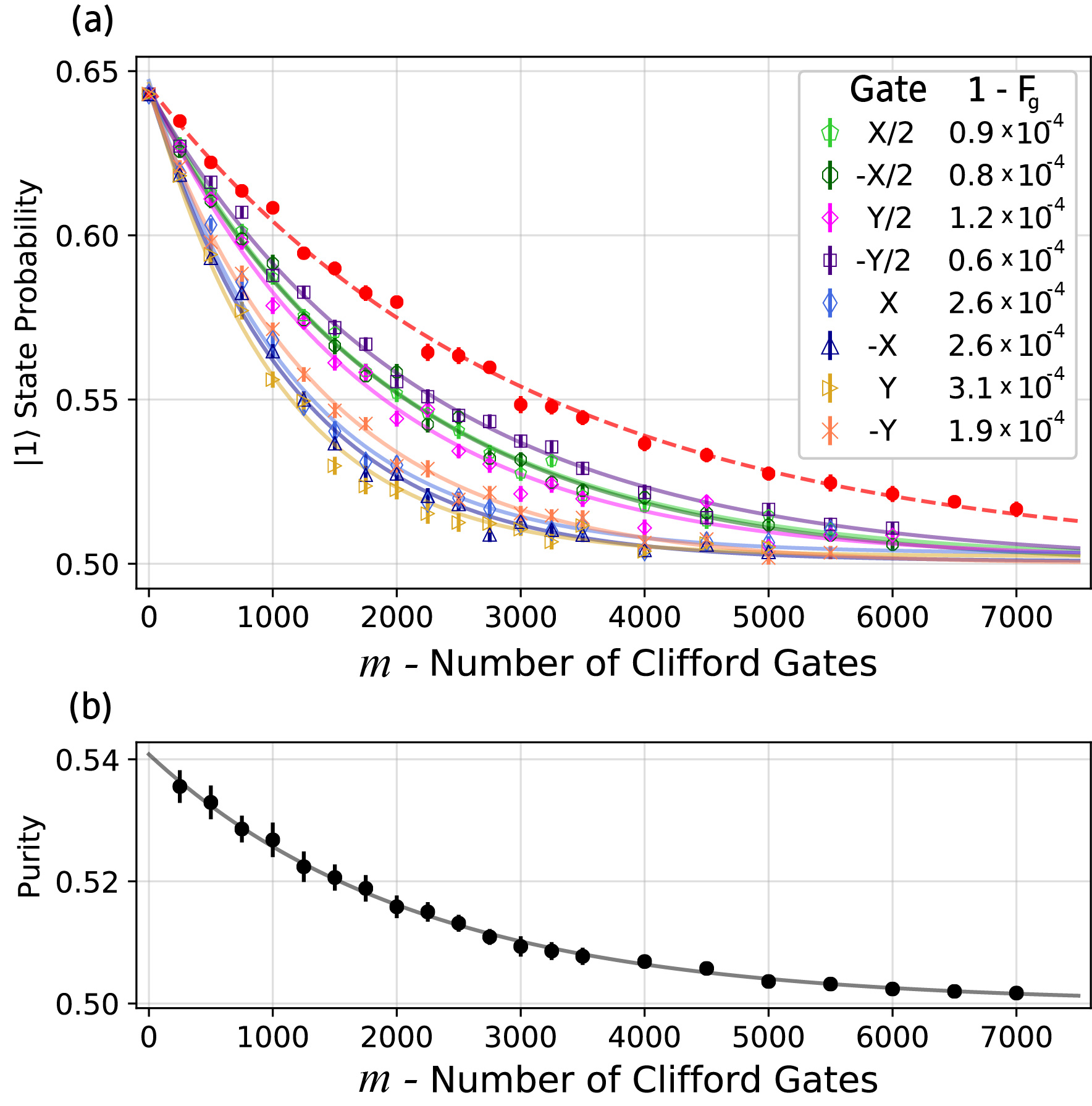}
	\caption{(a) Randomized Benchmarking (RB). The red curve is the reference RB with an average Clifford gate error rate of $(1.7 \pm 0.2) \times 10^{-4}$. The average gate fidelity of the physical gates used to generate the Clifford group is 99.991(1) \%. The colored curves are interleaved RB, each color encode a given interleaved gate. The relative error on the gate errors given in the caption is of the order of 10\%. (b) Purity Benchmarking (PB). Decay of the purity of the quantum state versus number of Cliffords. The gate error due to decoherence is $r_{\textup{dec, gate}}= 0.6 \times 10^{-4}$ establishing an upper bound on the achievable average gate fidelity of 99.994 \%. All the curves are averaged over 50 random realizations of the experiment.}
	\label{fig:Fig3}
\end{figure}

Coherent control over a single qubit can be more thoroughly characterized using the randomized benchmarking (RB) technique \cite{Magesan2011ScalableProcesses,Magesan2012EfficientBenchmarking}. In a RB sequence, $m$ randomly chosen Clifford gates are performed on the qubit before applying a single recovery gate, aimed at bringing the state vector back to the initial state. In the absence of gate errors, the entire random sequence amounts to identity operation. In reality, the excited state probability $\textup{p}(|1 \rangle)$ decays with the sequence length $m$ as $A + Bp^m$, where $p$ is the depolarization parameter, and $A, B$ are constants that absorb state preparation and measurement (SPAM) errors (see the red curve in Fig.~\ref{fig:Fig3}a). We extract an average error rate of a Clifford operation $r_{\textup{cliff}}$ given by $r_{\textup{cliff}} = (1-p)/2 = (1.7 \pm 0.2) \times 10^{-4}$. Because each Clifford operation is composed on average of 1.833 physical gates (we do not count the identity gate), the average physical gate fidelity is given by $F_{\textup{g, avg}} = 1 - r_{\textup{cliff}}/1.833 = 0.99991(1)$. To our knowledge, a higher fidelity number has been possible only in refined trapped ion demonstrations \cite{Bermudez2017AssessingComputation}.

The fidelity of each physical gate in the list ($\pm X, \pm Y, \pm X/2, \pm Y/2$) can be extracted using an interleaved RB sequence. The sequence is the same as the standard RB sequence, except now a given gate is interleaved between each Clifford operation. The resulting curve follows the same decay profile as for the standard RB, but with a depolarization parameter $p_{\textup{gate}}$. The physical gate error is given by $r_{\textup{gate}} =(1-p_{\textup{gate}}/p)/2 = 1-F_{\textup{g}}$, where $p$ is the depolarization parameter obtained from reference RB (Fig.~\ref{fig:Fig3}a, inset). The decoherence contribution to the gate error can be estimated using the purity benchmarking (PB) procedure~\cite{Feng2016EstimatingQubit, Wallman2015EstimatingNoise,Chen2018MetrologyQubits}. Purity benchmarking consists of performing state tomography of the qubit at the end of the RB sequence instead of the recovery gate. The purity $P = \textup{tr}(\rho^2)$ of the qubit state decays as  $A' + B'u^{m-1}$ (see Fig.~\ref{fig:Fig3}b), where $u$ is called the unitary and $A', B'$ are constants. The error rate due to decoherence per Clifford gate is $r_{\textup{dec, Cliff}} = (1-\sqrt{u})/2 \simeq 1.1 \times 10^{-4}$, and the error rate due to decoherence per gate is $r_{\textup{dec, gate}} = r_{\textup{dec, Cliff}}/1.833 \simeq 0.6 \times 10^{-4}$. We thus conclude that most of the gate error is caused by incoherent processes and hence can be reduced even further by shortening the pulses.

To pinpoint possible decoherence mechanisms, we first examine the single-shot readout data. Taking advantage of the clear bifurcation of the cavity transmission signal in the IQ-plane (see Supplementary Note 2A), we determine the qubit-cavity dispersive shift $\chi_{01}/2\pi = 1.3~\textrm{MHz}$. We note that such an interaction rate is typical to transmon-based circuit quantum electrodynamics. Given the value of $\chi_{01}$, the pure dephasing time $1/(1/T_2^* - 1/2T_1) \approx 4.5~\textrm{ms}$ can be explained by the presence of approximately $4\times 10^{-4}$ photons on average in the cavity. This number corresponds to a cavity mode temperature of about $45~\textrm{mK}$, which can be improved further with better thermalization of measurement lines~\cite{Yeh2017MicrowaveMK,Wang2019CavityQubits,Schuster2005AcField}. The single-shot historgrams also provide an accurate estimate of the qubit temperature, extracted from fitting the equilibrium populations of states $|0\rangle$ and $|1\rangle$ (see Supplementary Note 2A). We find the qubit temperature of $25~\textrm{mK}$, about twice lower than typical values reported in the case of transmons \cite{Jin2015ThermalQubit}. Increasing the refrigerator temperature from the base value under $10~\textrm{mK}$ to $25~\textrm{mK}$ did not modify $T_1$ appreciably, but heating to $50~\textrm{mK}$ increased the relaxation rate by a factor of three, in agreement with the stimulated emission factor. Thus, our circuit and setup appear relatively well-thermalized, and energy relaxation is the major limitation to coherence. 

\begin{figure}[t]
	\centering
	\includegraphics[width=\linewidth]{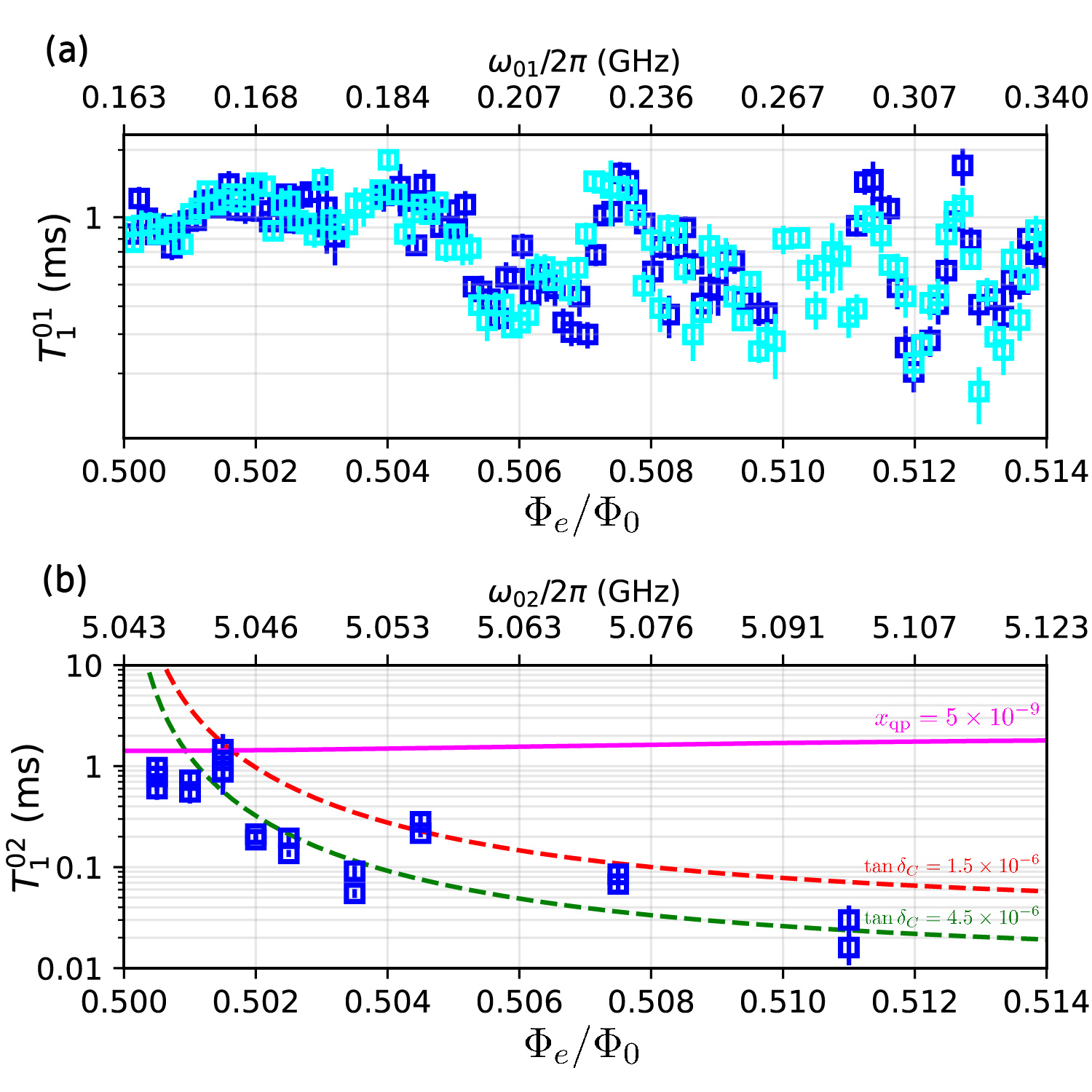}
    \caption{(a) Energy relaxation time $T_1^{01}$ of the $|0\rangle - |1\rangle$  transition versus external magnetic flux. Marker color indicates two different scans taken 24 hours apart. (b) Energy relaxation time $T_1^{02}$ of the $|0\rangle - |2\rangle$ transition versus external magnetic flux. Red and green dotted curves are the limits imposed by dielectric loss with loss tangents of $1.5$ and $4.5 \times 10^{-6}$, respectively. The pink curve is the limit imposed by quasiparticles tunneling across the small Josephson junction with an effective reduced quasiparticle density $x_{\textrm{qp}} = 5 \times 10^{-9}$.}
	\label{fig:Fig4}
\end{figure}

Two additional tests were performed to gain insights into energy relaxation mechanisms. First, the measurement of the qubit relaxation time $T_1\equiv T_1^{01}$ in Fig.~\ref{fig:Fig2} was repeated as a function of flux bias near the half-integer flux quantum sweet spot (Fig.~\ref{fig:Fig4}a). In a more elaborate experiment, we measured the rate $1/T_1^{02}$ of direct relaxation between states $|2\rangle$ and $|0\rangle$ (Fig.~\ref{fig:Fig4}b). This process should not be confused with an indirect relaxation via state $|1\rangle$, i.e. involving a rapid decay $|2\rangle \rightarrow |1\rangle$, which has a characteristic time $T_1^{12} \approx 10-20~\mu\textrm{s}$ (see Supplementary Note 6). Our protocol for measuring $T_1^{02}$ (Supplementary Note 7) requires a large ratio of $T_1^{01}/T_1^{12} \sim 10^2$, as well as an accurate calibration of the qubit temperature and lifetime $T_1^{01}$ at every flux bias. Namely, we apply a saturating Rabi drive of a given duration $\tau$ to the $|1\rangle - |2\rangle$ transition, wait for a period of a few times $T_1^{12}$, and then record the change in the population $p_0(\tau)$ of state $|0\rangle$. The deviation of $p_0$ from its equilibrium value encodes the transfer of population from state $|2\rangle$ to state $|0\rangle$. We model the quantity $p_0(\tau)$ using a 3-level optical pumping scheme and extract the value of $T_1^{02}$ as the sole adjustable parameter.

The data in Fig.~\ref{fig:Fig4}a differ from the results of an earlier experiment on a fluxonium qubit, which reported the observation of phase-sensitive dissipation caused by quasiparticle tunneling \cite{Pop2014CoherentQuasiparticles}. Namely, we observe no peak in $T_1^{01}$ at the half-integer flux bias. Furthermore, the relaxation signal shows no double-exponential character, and the decay time varies in a seemingly random but reproducible fashion with the qubit frequency. Such behavior of $T_1^{01}$ versus flux rules out qubit relaxation by quasiparticle tunneling and rather points at absorption by material defects \cite{Klimov2018FluctuationsQubits} as the main source of dissipation. According to theory   \cite{Catelani2011RelaxationQubits} (see Supplementary Note 5), the measured value $T_1^{01} > 1~\textrm{ms}$ at the half-integer flux bias imposes an upper bound on the normalized quasiparticle density, $x_{\textrm{qp}} < 6 \times 10^{-10}$. This number corresponds to having no more than one resident quasiparticle, on average, in the entire circuit (Fig.~\ref{fig:Fig1}a).

The relaxation time $T_1^{02}$ rapidly grows as the flux bias approaches the half-integer value (Fig.~4b), where the $|0\rangle - |2\rangle$ transition ($\omega_{02}/2\pi \approx 5~\textrm{GHz}$) is dipole-forbidden by the parity selection rule. The data in Fig.~\ref{fig:Fig4}b agree with the dielectric loss model, $1/T_1^{20} = 32\pi (E_C/h)|\langle 0|\hat{n}|2\rangle|^2\tan\delta_C$ (see Supplementary Note 4), where the loss tangent $\tan\delta_C$ of the total capacitance $C$ across the weak junction ($E_C = e^2/2C$) belongs to a range $\tan\delta_C \approx (1.5-4.5)\times 10^{-6}$. These values of $\tan\delta_C$ are several times larger than those reported for best transmon qubits \cite{Wang2015SurfaceQubits}, which most likely reflects our sub-optimal fabrication procedures and antenna geometry.

Interestingly, the value of $T_1^{02}$ saturates at about $T_1^{02} \approx 1.5~\textrm{ms}$, suggesting the presence of an apparently parity-violating decay process. This saturation can be explained by a thermal excitation $|2\rangle \rightarrow |3\rangle$, the estimated rate of which is indeed around $1~\textrm{ms}$ ($\omega_{23}/2\pi = 1.66~\textrm{GHz}$, $T = 25~\textrm{mK}$), followed by a rapid direct relaxation $|3\rangle \rightarrow |0\rangle$ (see Supplementary Note 7D). More intriguingly, though, theory \cite{Glazman2020BogoliubovQubits} does predict a parity-violating direct decay $|2\rangle \rightarrow |0\rangle$ due to the quasiparticle tunneling across the fluxonium's weak junction, with a rate governed by the matrix element $\langle 2|\cos (\hat{\varphi}/2)|0\rangle$. Even in the absence of resident quasiparticles ($x_{\textrm{qp}} = 0$), the decay may occur via a photon-assisted tunneling (PAT) process \cite{Glazman2020BogoliubovQubits, Houzet2019Photon-assistedQubit}: the weak junction first absorbs a stray photon at an energy above the superconducting gap, a Cooper pair consequently splits into two quasiparticles at the opposite sides of the junction, and a tunneling event follows. Note, the PAT process may not be efficient for the chain junctions because of their physically small leads. The intensity of stray radiation can be characterized by an apparent quasiparticle density $x_{\textrm{qp}}^{\textrm{PAT}}$, which adds to the resident quasiparticle density, i.e. $x_{\textrm{qp}}\rightarrow x_{\textrm{qp}} + x_{\textrm{qp}}^{\textrm{PAT}}$ and our observation that $T_1^{02} > 1.5~\textrm{ms}$ translates into a direct upper bound $x_{\textrm{qp}} < 5\times 10^{-9}$. For a transmon qubit, our bound corresponds to quasiparticle-induced lifetime longer than $4~\textrm{ms}$.

Our data challenge recent projections that quasiparticle production due to the background radioactivity or cosmic rays poses a serious threat to coherence of superconducting qubits \cite{Martinis2020SavingRays,Vepsalainen2020ImpactCoherence}. Instead, we conclude that reducing the loss due to material defects~\cite{Place2020NewMilliseconds} and improving the measurement line thermalization \cite{Wang2019CavityQubits} would push coherence well into the millisecond range. In the meantime, our demonstration of a stable Ramsey coherence time $T_2^* > 1~\textrm{ms}$ and an average gate fidelity $F_{\mathrm{g,avg}} > 0.9999$ establishes a precedent for the most coherent superconducting qubit and the most controllable engineered quantum system to date.


\begin{thebibliography}{49}%
\makeatletter
\providecommand \@ifxundefined [1]{%
 \@ifx{#1\undefined}
}%
\providecommand \@ifnum [1]{%
 \ifnum #1\expandafter \@firstoftwo
 \else \expandafter \@secondoftwo
 \fi
}%
\providecommand \@ifx [1]{%
 \ifx #1\expandafter \@firstoftwo
 \else \expandafter \@secondoftwo
 \fi
}%
\providecommand \natexlab [1]{#1}%
\providecommand \enquote  [1]{``#1''}%
\providecommand \bibnamefont  [1]{#1}%
\providecommand \bibfnamefont [1]{#1}%
\providecommand \citenamefont [1]{#1}%
\providecommand \href@noop [0]{\@secondoftwo}%
\providecommand \href [0]{\begingroup \@sanitize@url \@href}%
\providecommand \@href[1]{\@@startlink{#1}\@@href}%
\providecommand \@@href[1]{\endgroup#1\@@endlink}%
\providecommand \@sanitize@url [0]{\catcode `\\12\catcode `\$12\catcode
  `\&12\catcode `\#12\catcode `\^12\catcode `\_12\catcode `\%12\relax}%
\providecommand \@@startlink[1]{}%
\providecommand \@@endlink[0]{}%
\providecommand \url  [0]{\begingroup\@sanitize@url \@url }%
\providecommand \@url [1]{\endgroup\@href {#1}{\urlprefix }}%
\providecommand \urlprefix  [0]{URL }%
\providecommand \Eprint [0]{\href }%
\providecommand \doibase [0]{https://doi.org/}%
\providecommand \selectlanguage [0]{\@gobble}%
\providecommand \bibinfo  [0]{\@secondoftwo}%
\providecommand \bibfield  [0]{\@secondoftwo}%
\providecommand \translation [1]{[#1]}%
\providecommand \BibitemOpen [0]{}%
\providecommand \bibitemStop [0]{}%
\providecommand \bibitemNoStop [0]{.\EOS\space}%
\providecommand \EOS [0]{\spacefactor3000\relax}%
\providecommand \BibitemShut  [1]{\csname bibitem#1\endcsname}%
\let\auto@bib@innerbib\@empty
\bibitem [{\citenamefont {Devoret}\ and\ \citenamefont
  {Schoelkopf}(2013)}]{Devoret2013SuperconductingOutlook}%
  \BibitemOpen
  \bibfield  {author} {\bibinfo {author} {\bibfnamefont {M.~H.}\ \bibnamefont
  {Devoret}}\ and\ \bibinfo {author} {\bibfnamefont {R.~J.}\ \bibnamefont
  {Schoelkopf}},\ }\href {https://doi.org/10.1126/science.1231930} {\bibinfo
  {title} {{Superconducting circuits for quantum information: An outlook}}}
  (\bibinfo {year} {2013})\BibitemShut {NoStop}%
\bibitem [{\citenamefont {Kjaergaard}\ \emph {et~al.}(2020)\citenamefont
  {Kjaergaard}, \citenamefont {Schwartz}, \citenamefont {Braum{\"{u}}ller},
  \citenamefont {Krantz}, \citenamefont {Wang}, \citenamefont {Gustavsson},\
  and\ \citenamefont {Oliver}}]{Kjaergaard2020SuperconductingPlay}%
  \BibitemOpen
  \bibfield  {author} {\bibinfo {author} {\bibfnamefont {M.}~\bibnamefont
  {Kjaergaard}}, \bibinfo {author} {\bibfnamefont {M.~E.}\ \bibnamefont
  {Schwartz}}, \bibinfo {author} {\bibfnamefont {J.}~\bibnamefont
  {Braum{\"{u}}ller}}, \bibinfo {author} {\bibfnamefont {P.}~\bibnamefont
  {Krantz}}, \bibinfo {author} {\bibfnamefont {J.~I.-J.}\ \bibnamefont {Wang}},
  \bibinfo {author} {\bibfnamefont {S.}~\bibnamefont {Gustavsson}},\ and\
  \bibinfo {author} {\bibfnamefont {W.~D.}\ \bibnamefont {Oliver}},\ }\href
  {https://doi.org/10.1146/annurev-conmatphys-031119-050605} {\bibinfo {title}
  {{Superconducting Qubits: Current State of Play}}} (\bibinfo {year}
  {2020})\BibitemShut {NoStop}%
\bibitem [{\citenamefont {Nakamura}\ \emph {et~al.}(1999)\citenamefont
  {Nakamura}, \citenamefont {Pashkin},\ and\ \citenamefont
  {Tsai}}]{Nakamura1999CoherentBox}%
  \BibitemOpen
  \bibfield  {author} {\bibinfo {author} {\bibfnamefont {Y.}~\bibnamefont
  {Nakamura}}, \bibinfo {author} {\bibfnamefont {Y.~A.}\ \bibnamefont
  {Pashkin}},\ and\ \bibinfo {author} {\bibfnamefont {J.~S.}\ \bibnamefont
  {Tsai}},\ }\bibfield  {title} {\bibinfo {title} {{Coherent control of
  macroscopic quantum states in a single-Cooper-pair box}},\ }\href
  {https://doi.org/10.1038/19718} {\bibfield  {journal} {\bibinfo  {journal}
  {Nature}\ }\textbf {\bibinfo {volume} {398}},\ \bibinfo {pages} {786}
  (\bibinfo {year} {1999})}\BibitemShut {NoStop}%
\bibitem [{\citenamefont {Vion}\ \emph {et~al.}(2002)\citenamefont {Vion},
  \citenamefont {Aassime}, \citenamefont {Cottet}, \citenamefont {Joyez},
  \citenamefont {Pothier}, \citenamefont {Urbina}, \citenamefont {Esteve},\
  and\ \citenamefont {Devoret}}]{Vion2002ManipulatingCircuit}%
  \BibitemOpen
  \bibfield  {author} {\bibinfo {author} {\bibfnamefont {D.}~\bibnamefont
  {Vion}}, \bibinfo {author} {\bibfnamefont {A.}~\bibnamefont {Aassime}},
  \bibinfo {author} {\bibfnamefont {A.}~\bibnamefont {Cottet}}, \bibinfo
  {author} {\bibfnamefont {P.}~\bibnamefont {Joyez}}, \bibinfo {author}
  {\bibfnamefont {H.}~\bibnamefont {Pothier}}, \bibinfo {author} {\bibfnamefont
  {C.}~\bibnamefont {Urbina}}, \bibinfo {author} {\bibfnamefont
  {D.}~\bibnamefont {Esteve}},\ and\ \bibinfo {author} {\bibfnamefont {M.~H.}\
  \bibnamefont {Devoret}},\ }\bibfield  {title} {\bibinfo {title}
  {{Manipulating the quantum state of an electrical circuit}},\ }\href
  {https://doi.org/10.1126/science.1069372} {\bibfield  {journal} {\bibinfo
  {journal} {Science}\ }\textbf {\bibinfo {volume} {296}},\ \bibinfo {pages}
  {886} (\bibinfo {year} {2002})}\BibitemShut {NoStop}%
\bibitem [{\citenamefont {Paik}\ \emph {et~al.}(2011)\citenamefont {Paik},
  \citenamefont {Schuster}, \citenamefont {Bishop}, \citenamefont {Kirchmair},
  \citenamefont {Catelani}, \citenamefont {Sears}, \citenamefont {Johnson},
  \citenamefont {Reagor}, \citenamefont {Frunzio}, \citenamefont {Glazman},
  \citenamefont {Girvin}, \citenamefont {Devoret},\ and\ \citenamefont
  {Schoelkopf}}]{Paik2011ObservationArchitecture}%
  \BibitemOpen
  \bibfield  {author} {\bibinfo {author} {\bibfnamefont {H.}~\bibnamefont
  {Paik}}, \bibinfo {author} {\bibfnamefont {D.~I.}\ \bibnamefont {Schuster}},
  \bibinfo {author} {\bibfnamefont {L.~S.}\ \bibnamefont {Bishop}}, \bibinfo
  {author} {\bibfnamefont {G.}~\bibnamefont {Kirchmair}}, \bibinfo {author}
  {\bibfnamefont {G.}~\bibnamefont {Catelani}}, \bibinfo {author}
  {\bibfnamefont {A.~P.}\ \bibnamefont {Sears}}, \bibinfo {author}
  {\bibfnamefont {B.~R.}\ \bibnamefont {Johnson}}, \bibinfo {author}
  {\bibfnamefont {M.~J.}\ \bibnamefont {Reagor}}, \bibinfo {author}
  {\bibfnamefont {L.}~\bibnamefont {Frunzio}}, \bibinfo {author} {\bibfnamefont
  {L.~I.}\ \bibnamefont {Glazman}}, \bibinfo {author} {\bibfnamefont {S.~M.}\
  \bibnamefont {Girvin}}, \bibinfo {author} {\bibfnamefont {M.~H.}\
  \bibnamefont {Devoret}},\ and\ \bibinfo {author} {\bibfnamefont {R.~J.}\
  \bibnamefont {Schoelkopf}},\ }\bibfield  {title} {\bibinfo {title}
  {{Observation of high coherence in Josephson junction qubits measured in a
  three-dimensional circuit QED architecture}},\ }\href
  {https://doi.org/10.1103/PhysRevLett.107.240501} {\bibfield  {journal}
  {\bibinfo  {journal} {Physical Review Letters}\ }\textbf {\bibinfo {volume}
  {107}},\ \bibinfo {pages} {240501} (\bibinfo {year} {2011})}\BibitemShut
  {NoStop}%
\bibitem [{\citenamefont {Barends}\ \emph {et~al.}(2013)\citenamefont
  {Barends}, \citenamefont {Kelly}, \citenamefont {Megrant}, \citenamefont
  {Sank}, \citenamefont {Jeffrey}, \citenamefont {Chen}, \citenamefont {Yin},
  \citenamefont {Chiaro}, \citenamefont {Mutus}, \citenamefont {Neill},
  \citenamefont {O'Malley}, \citenamefont {Roushan}, \citenamefont {Wenner},
  \citenamefont {White}, \citenamefont {Cleland},\ and\ \citenamefont
  {Martinis}}]{Barends2013CoherentCircuits}%
  \BibitemOpen
  \bibfield  {author} {\bibinfo {author} {\bibfnamefont {R.}~\bibnamefont
  {Barends}}, \bibinfo {author} {\bibfnamefont {J.}~\bibnamefont {Kelly}},
  \bibinfo {author} {\bibfnamefont {A.}~\bibnamefont {Megrant}}, \bibinfo
  {author} {\bibfnamefont {D.}~\bibnamefont {Sank}}, \bibinfo {author}
  {\bibfnamefont {E.}~\bibnamefont {Jeffrey}}, \bibinfo {author} {\bibfnamefont
  {Y.}~\bibnamefont {Chen}}, \bibinfo {author} {\bibfnamefont {Y.}~\bibnamefont
  {Yin}}, \bibinfo {author} {\bibfnamefont {B.}~\bibnamefont {Chiaro}},
  \bibinfo {author} {\bibfnamefont {J.}~\bibnamefont {Mutus}}, \bibinfo
  {author} {\bibfnamefont {C.}~\bibnamefont {Neill}}, \bibinfo {author}
  {\bibfnamefont {P.}~\bibnamefont {O'Malley}}, \bibinfo {author}
  {\bibfnamefont {P.}~\bibnamefont {Roushan}}, \bibinfo {author} {\bibfnamefont
  {J.}~\bibnamefont {Wenner}}, \bibinfo {author} {\bibfnamefont {T.~C.}\
  \bibnamefont {White}}, \bibinfo {author} {\bibfnamefont {A.~N.}\ \bibnamefont
  {Cleland}},\ and\ \bibinfo {author} {\bibfnamefont {J.~M.}\ \bibnamefont
  {Martinis}},\ }\bibfield  {title} {\bibinfo {title} {{Coherent josephson
  qubit suitable for scalable quantum integrated circuits}},\ }\bibfield
  {journal} {\bibinfo  {journal} {Physical Review Letters}\ }\textbf {\bibinfo
  {volume} {111}},\ \href {https://doi.org/10.1103/PhysRevLett.111.080502}
  {10.1103/PhysRevLett.111.080502} (\bibinfo {year} {2013})\BibitemShut
  {NoStop}%
\bibitem [{\citenamefont {Yan}\ \emph {et~al.}(2016)\citenamefont {Yan},
  \citenamefont {Gustavsson}, \citenamefont {Kamal}, \citenamefont {Birenbaum},
  \citenamefont {Sears}, \citenamefont {Hover}, \citenamefont {Gudmundsen},
  \citenamefont {Rosenberg}, \citenamefont {Samach}, \citenamefont {Weber},
  \citenamefont {Yoder}, \citenamefont {Orlando}, \citenamefont {Clarke},
  \citenamefont {Kerman},\ and\ \citenamefont
  {Oliver}}]{Yan2016TheReproducibility}%
  \BibitemOpen
  \bibfield  {author} {\bibinfo {author} {\bibfnamefont {F.}~\bibnamefont
  {Yan}}, \bibinfo {author} {\bibfnamefont {S.}~\bibnamefont {Gustavsson}},
  \bibinfo {author} {\bibfnamefont {A.}~\bibnamefont {Kamal}}, \bibinfo
  {author} {\bibfnamefont {J.}~\bibnamefont {Birenbaum}}, \bibinfo {author}
  {\bibfnamefont {A.~P.}\ \bibnamefont {Sears}}, \bibinfo {author}
  {\bibfnamefont {D.}~\bibnamefont {Hover}}, \bibinfo {author} {\bibfnamefont
  {T.~J.}\ \bibnamefont {Gudmundsen}}, \bibinfo {author} {\bibfnamefont
  {D.}~\bibnamefont {Rosenberg}}, \bibinfo {author} {\bibfnamefont
  {G.}~\bibnamefont {Samach}}, \bibinfo {author} {\bibfnamefont
  {S.}~\bibnamefont {Weber}}, \bibinfo {author} {\bibfnamefont {J.~L.}\
  \bibnamefont {Yoder}}, \bibinfo {author} {\bibfnamefont {T.~P.}\ \bibnamefont
  {Orlando}}, \bibinfo {author} {\bibfnamefont {J.}~\bibnamefont {Clarke}},
  \bibinfo {author} {\bibfnamefont {A.~J.}\ \bibnamefont {Kerman}},\ and\
  \bibinfo {author} {\bibfnamefont {W.~D.}\ \bibnamefont {Oliver}},\ }\bibfield
   {title} {\bibinfo {title} {{The flux qubit revisited to enhance coherence
  and reproducibility}},\ }\bibfield  {journal} {\bibinfo  {journal} {Nature
  Communications}\ }\textbf {\bibinfo {volume} {7}},\ \href
  {https://doi.org/10.1038/ncomms12964} {10.1038/ncomms12964} (\bibinfo {year}
  {2016})\BibitemShut {NoStop}%
\bibitem [{\citenamefont {Place}\ \emph {et~al.}(2020)\citenamefont {Place},
  \citenamefont {Rodgers}, \citenamefont {Mundada}, \citenamefont {Smitham},
  \citenamefont {Fitzpatrick}, \citenamefont {Leng}, \citenamefont {Premkumar},
  \citenamefont {Bryon}, \citenamefont {Sussman}, \citenamefont {Cheng},
  \citenamefont {Madhavan}, \citenamefont {Babla}, \citenamefont {J{\"{a}}ck},
  \citenamefont {Gyenis}, \citenamefont {Yao}, \citenamefont {Cava},
  \citenamefont {de~Leon},\ and\ \citenamefont
  {Houck}}]{Place2020NewMilliseconds}%
  \BibitemOpen
  \bibfield  {author} {\bibinfo {author} {\bibfnamefont {A.~P.}\ \bibnamefont
  {Place}}, \bibinfo {author} {\bibfnamefont {L.~V.}\ \bibnamefont {Rodgers}},
  \bibinfo {author} {\bibfnamefont {P.}~\bibnamefont {Mundada}}, \bibinfo
  {author} {\bibfnamefont {B.~M.}\ \bibnamefont {Smitham}}, \bibinfo {author}
  {\bibfnamefont {M.}~\bibnamefont {Fitzpatrick}}, \bibinfo {author}
  {\bibfnamefont {Z.}~\bibnamefont {Leng}}, \bibinfo {author} {\bibfnamefont
  {A.}~\bibnamefont {Premkumar}}, \bibinfo {author} {\bibfnamefont
  {J.}~\bibnamefont {Bryon}}, \bibinfo {author} {\bibfnamefont
  {S.}~\bibnamefont {Sussman}}, \bibinfo {author} {\bibfnamefont
  {G.}~\bibnamefont {Cheng}}, \bibinfo {author} {\bibfnamefont
  {T.}~\bibnamefont {Madhavan}}, \bibinfo {author} {\bibfnamefont {H.~K.}\
  \bibnamefont {Babla}}, \bibinfo {author} {\bibfnamefont {B.}~\bibnamefont
  {J{\"{a}}ck}}, \bibinfo {author} {\bibfnamefont {A.}~\bibnamefont {Gyenis}},
  \bibinfo {author} {\bibfnamefont {N.}~\bibnamefont {Yao}}, \bibinfo {author}
  {\bibfnamefont {R.~J.}\ \bibnamefont {Cava}}, \bibinfo {author}
  {\bibfnamefont {N.~P.}\ \bibnamefont {de~Leon}},\ and\ \bibinfo {author}
  {\bibfnamefont {A.~A.}\ \bibnamefont {Houck}},\ }\bibfield  {title} {\bibinfo
  {title} {{New material platform for superconducting transmon qubits with
  coherence times exceeding 0.3 milliseconds}},\ }\href@noop {} {\bibfield
  {journal} {\bibinfo  {journal} {arXiv:2003.00024}\ } (\bibinfo {year}
  {2020})}\BibitemShut {NoStop}%
\bibitem [{\citenamefont {Rigetti}\ \emph {et~al.}(2012)\citenamefont
  {Rigetti}, \citenamefont {Gambetta}, \citenamefont {Poletto}, \citenamefont
  {Plourde}, \citenamefont {Chow}, \citenamefont {C{\'{o}}rcoles},
  \citenamefont {Smolin}, \citenamefont {Merkel}, \citenamefont {Rozen},
  \citenamefont {Keefe}, \citenamefont {Rothwell}, \citenamefont {Ketchen},\
  and\ \citenamefont {Steffen}}]{Rigetti2012SuperconductingMs}%
  \BibitemOpen
  \bibfield  {author} {\bibinfo {author} {\bibfnamefont {C.}~\bibnamefont
  {Rigetti}}, \bibinfo {author} {\bibfnamefont {J.~M.}\ \bibnamefont
  {Gambetta}}, \bibinfo {author} {\bibfnamefont {S.}~\bibnamefont {Poletto}},
  \bibinfo {author} {\bibfnamefont {B.~L.}\ \bibnamefont {Plourde}}, \bibinfo
  {author} {\bibfnamefont {J.~M.}\ \bibnamefont {Chow}}, \bibinfo {author}
  {\bibfnamefont {A.~D.}\ \bibnamefont {C{\'{o}}rcoles}}, \bibinfo {author}
  {\bibfnamefont {J.~A.}\ \bibnamefont {Smolin}}, \bibinfo {author}
  {\bibfnamefont {S.~T.}\ \bibnamefont {Merkel}}, \bibinfo {author}
  {\bibfnamefont {J.~R.}\ \bibnamefont {Rozen}}, \bibinfo {author}
  {\bibfnamefont {G.~A.}\ \bibnamefont {Keefe}}, \bibinfo {author}
  {\bibfnamefont {M.~B.}\ \bibnamefont {Rothwell}}, \bibinfo {author}
  {\bibfnamefont {M.~B.}\ \bibnamefont {Ketchen}},\ and\ \bibinfo {author}
  {\bibfnamefont {M.}~\bibnamefont {Steffen}},\ }\bibfield  {title} {\bibinfo
  {title} {{Superconducting qubit in a waveguide cavity with a coherence time
  approaching 0.1 ms}},\ }\bibfield  {journal} {\bibinfo  {journal} {Physical
  Review B - Condensed Matter and Materials Physics}\ }\textbf {\bibinfo
  {volume} {86}},\ \href {https://doi.org/10.1103/PhysRevB.86.100506}
  {10.1103/PhysRevB.86.100506} (\bibinfo {year} {2012})\BibitemShut {NoStop}%
\bibitem [{\citenamefont {Arute}\ \emph {et~al.}(2019)\citenamefont {Arute},
  \citenamefont {Arya}, \citenamefont {Babbush}, \citenamefont {Bacon},
  \citenamefont {Bardin}, \citenamefont {Barends}, \citenamefont {Biswas},
  \citenamefont {Boixo}, \citenamefont {Brandao}, \citenamefont {Buell},
  \citenamefont {Burkett}, \citenamefont {Chen}, \citenamefont {Chen},
  \citenamefont {Chiaro}, \citenamefont {Collins}, \citenamefont {Courtney},
  \citenamefont {Dunsworth}, \citenamefont {Farhi}, \citenamefont {Foxen},
  \citenamefont {Fowler}, \citenamefont {Gidney}, \citenamefont {Giustina},
  \citenamefont {Graff}, \citenamefont {Guerin}, \citenamefont {Habegger},
  \citenamefont {Harrigan}, \citenamefont {Hartmann}, \citenamefont {Ho},
  \citenamefont {Hoffmann}, \citenamefont {Huang}, \citenamefont {Humble},
  \citenamefont {Isakov}, \citenamefont {Jeffrey}, \citenamefont {Jiang},
  \citenamefont {Kafri}, \citenamefont {Kechedzhi}, \citenamefont {Kelly},
  \citenamefont {Klimov}, \citenamefont {Knysh}, \citenamefont {Korotkov},
  \citenamefont {Kostritsa}, \citenamefont {Landhuis}, \citenamefont
  {Lindmark}, \citenamefont {Lucero}, \citenamefont {Lyakh}, \citenamefont
  {Mandr{\`{a}}}, \citenamefont {McClean}, \citenamefont {McEwen},
  \citenamefont {Megrant}, \citenamefont {Mi}, \citenamefont {Michielsen},
  \citenamefont {Mohseni}, \citenamefont {Mutus}, \citenamefont {Naaman},
  \citenamefont {Neeley}, \citenamefont {Neill}, \citenamefont {Niu},
  \citenamefont {Ostby}, \citenamefont {Petukhov}, \citenamefont {Platt},
  \citenamefont {Quintana}, \citenamefont {Rieffel}, \citenamefont {Roushan},
  \citenamefont {Rubin}, \citenamefont {Sank}, \citenamefont {Satzinger},
  \citenamefont {Smelyanskiy}, \citenamefont {Sung}, \citenamefont
  {Trevithick}, \citenamefont {Vainsencher}, \citenamefont {Villalonga},
  \citenamefont {White}, \citenamefont {Yao}, \citenamefont {Yeh},
  \citenamefont {Zalcman}, \citenamefont {Neven},\ and\ \citenamefont
  {Martinis}}]{Arute2019QuantumProcessor}%
  \BibitemOpen
  \bibfield  {author} {\bibinfo {author} {\bibfnamefont {F.}~\bibnamefont
  {Arute}}, \bibinfo {author} {\bibfnamefont {K.}~\bibnamefont {Arya}},
  \bibinfo {author} {\bibfnamefont {R.}~\bibnamefont {Babbush}}, \bibinfo
  {author} {\bibfnamefont {D.}~\bibnamefont {Bacon}}, \bibinfo {author}
  {\bibfnamefont {J.~C.}\ \bibnamefont {Bardin}}, \bibinfo {author}
  {\bibfnamefont {R.}~\bibnamefont {Barends}}, \bibinfo {author} {\bibfnamefont
  {R.}~\bibnamefont {Biswas}}, \bibinfo {author} {\bibfnamefont
  {S.}~\bibnamefont {Boixo}}, \bibinfo {author} {\bibfnamefont {F.~G.}\
  \bibnamefont {Brandao}}, \bibinfo {author} {\bibfnamefont {D.~A.}\
  \bibnamefont {Buell}}, \bibinfo {author} {\bibfnamefont {B.}~\bibnamefont
  {Burkett}}, \bibinfo {author} {\bibfnamefont {Y.}~\bibnamefont {Chen}},
  \bibinfo {author} {\bibfnamefont {Z.}~\bibnamefont {Chen}}, \bibinfo {author}
  {\bibfnamefont {B.}~\bibnamefont {Chiaro}}, \bibinfo {author} {\bibfnamefont
  {R.}~\bibnamefont {Collins}}, \bibinfo {author} {\bibfnamefont
  {W.}~\bibnamefont {Courtney}}, \bibinfo {author} {\bibfnamefont
  {A.}~\bibnamefont {Dunsworth}}, \bibinfo {author} {\bibfnamefont
  {E.}~\bibnamefont {Farhi}}, \bibinfo {author} {\bibfnamefont
  {B.}~\bibnamefont {Foxen}}, \bibinfo {author} {\bibfnamefont
  {A.}~\bibnamefont {Fowler}}, \bibinfo {author} {\bibfnamefont
  {C.}~\bibnamefont {Gidney}}, \bibinfo {author} {\bibfnamefont
  {M.}~\bibnamefont {Giustina}}, \bibinfo {author} {\bibfnamefont
  {R.}~\bibnamefont {Graff}}, \bibinfo {author} {\bibfnamefont
  {K.}~\bibnamefont {Guerin}}, \bibinfo {author} {\bibfnamefont
  {S.}~\bibnamefont {Habegger}}, \bibinfo {author} {\bibfnamefont {M.~P.}\
  \bibnamefont {Harrigan}}, \bibinfo {author} {\bibfnamefont {M.~J.}\
  \bibnamefont {Hartmann}}, \bibinfo {author} {\bibfnamefont {A.}~\bibnamefont
  {Ho}}, \bibinfo {author} {\bibfnamefont {M.}~\bibnamefont {Hoffmann}},
  \bibinfo {author} {\bibfnamefont {T.}~\bibnamefont {Huang}}, \bibinfo
  {author} {\bibfnamefont {T.~S.}\ \bibnamefont {Humble}}, \bibinfo {author}
  {\bibfnamefont {S.~V.}\ \bibnamefont {Isakov}}, \bibinfo {author}
  {\bibfnamefont {E.}~\bibnamefont {Jeffrey}}, \bibinfo {author} {\bibfnamefont
  {Z.}~\bibnamefont {Jiang}}, \bibinfo {author} {\bibfnamefont
  {D.}~\bibnamefont {Kafri}}, \bibinfo {author} {\bibfnamefont
  {K.}~\bibnamefont {Kechedzhi}}, \bibinfo {author} {\bibfnamefont
  {J.}~\bibnamefont {Kelly}}, \bibinfo {author} {\bibfnamefont {P.~V.}\
  \bibnamefont {Klimov}}, \bibinfo {author} {\bibfnamefont {S.}~\bibnamefont
  {Knysh}}, \bibinfo {author} {\bibfnamefont {A.}~\bibnamefont {Korotkov}},
  \bibinfo {author} {\bibfnamefont {F.}~\bibnamefont {Kostritsa}}, \bibinfo
  {author} {\bibfnamefont {D.}~\bibnamefont {Landhuis}}, \bibinfo {author}
  {\bibfnamefont {M.}~\bibnamefont {Lindmark}}, \bibinfo {author}
  {\bibfnamefont {E.}~\bibnamefont {Lucero}}, \bibinfo {author} {\bibfnamefont
  {D.}~\bibnamefont {Lyakh}}, \bibinfo {author} {\bibfnamefont
  {S.}~\bibnamefont {Mandr{\`{a}}}}, \bibinfo {author} {\bibfnamefont {J.~R.}\
  \bibnamefont {McClean}}, \bibinfo {author} {\bibfnamefont {M.}~\bibnamefont
  {McEwen}}, \bibinfo {author} {\bibfnamefont {A.}~\bibnamefont {Megrant}},
  \bibinfo {author} {\bibfnamefont {X.}~\bibnamefont {Mi}}, \bibinfo {author}
  {\bibfnamefont {K.}~\bibnamefont {Michielsen}}, \bibinfo {author}
  {\bibfnamefont {M.}~\bibnamefont {Mohseni}}, \bibinfo {author} {\bibfnamefont
  {J.}~\bibnamefont {Mutus}}, \bibinfo {author} {\bibfnamefont
  {O.}~\bibnamefont {Naaman}}, \bibinfo {author} {\bibfnamefont
  {M.}~\bibnamefont {Neeley}}, \bibinfo {author} {\bibfnamefont
  {C.}~\bibnamefont {Neill}}, \bibinfo {author} {\bibfnamefont {M.~Y.}\
  \bibnamefont {Niu}}, \bibinfo {author} {\bibfnamefont {E.}~\bibnamefont
  {Ostby}}, \bibinfo {author} {\bibfnamefont {A.}~\bibnamefont {Petukhov}},
  \bibinfo {author} {\bibfnamefont {J.~C.}\ \bibnamefont {Platt}}, \bibinfo
  {author} {\bibfnamefont {C.}~\bibnamefont {Quintana}}, \bibinfo {author}
  {\bibfnamefont {E.~G.}\ \bibnamefont {Rieffel}}, \bibinfo {author}
  {\bibfnamefont {P.}~\bibnamefont {Roushan}}, \bibinfo {author} {\bibfnamefont
  {N.~C.}\ \bibnamefont {Rubin}}, \bibinfo {author} {\bibfnamefont
  {D.}~\bibnamefont {Sank}}, \bibinfo {author} {\bibfnamefont {K.~J.}\
  \bibnamefont {Satzinger}}, \bibinfo {author} {\bibfnamefont {V.}~\bibnamefont
  {Smelyanskiy}}, \bibinfo {author} {\bibfnamefont {K.~J.}\ \bibnamefont
  {Sung}}, \bibinfo {author} {\bibfnamefont {M.~D.}\ \bibnamefont
  {Trevithick}}, \bibinfo {author} {\bibfnamefont {A.}~\bibnamefont
  {Vainsencher}}, \bibinfo {author} {\bibfnamefont {B.}~\bibnamefont
  {Villalonga}}, \bibinfo {author} {\bibfnamefont {T.}~\bibnamefont {White}},
  \bibinfo {author} {\bibfnamefont {Z.~J.}\ \bibnamefont {Yao}}, \bibinfo
  {author} {\bibfnamefont {P.}~\bibnamefont {Yeh}}, \bibinfo {author}
  {\bibfnamefont {A.}~\bibnamefont {Zalcman}}, \bibinfo {author} {\bibfnamefont
  {H.}~\bibnamefont {Neven}},\ and\ \bibinfo {author} {\bibfnamefont {J.~M.}\
  \bibnamefont {Martinis}},\ }\bibfield  {title} {\bibinfo {title} {{Quantum
  supremacy using a programmable superconducting processor}},\ }\href
  {https://doi.org/10.1038/s41586-019-1666-5} {\bibfield  {journal} {\bibinfo
  {journal} {Nature}\ }\textbf {\bibinfo {volume} {574}},\ \bibinfo {pages}
  {505} (\bibinfo {year} {2019})}\BibitemShut {NoStop}%
\bibitem [{\citenamefont {Harrigan}\ \emph {et~al.}(2021)\citenamefont
  {Harrigan}, \citenamefont {Sung}, \citenamefont {Neeley}, \citenamefont
  {Satzinger}, \citenamefont {Arute}, \citenamefont {Arya}, \citenamefont
  {Atalaya}, \citenamefont {Bardin}, \citenamefont {Barends}, \citenamefont
  {Boixo}, \citenamefont {Broughton}, \citenamefont {Buckley}, \citenamefont
  {Buell}, \citenamefont {Burkett}, \citenamefont {Bushnell}, \citenamefont
  {Chen}, \citenamefont {Chen}, \citenamefont {{Ben Chiaro}}, \citenamefont
  {Collins}, \citenamefont {Courtney}, \citenamefont {Demura}, \citenamefont
  {Dunsworth}, \citenamefont {Eppens}, \citenamefont {Fowler}, \citenamefont
  {Foxen}, \citenamefont {Gidney}, \citenamefont {Giustina}, \citenamefont
  {Graff}, \citenamefont {Habegger}, \citenamefont {Ho}, \citenamefont {Hong},
  \citenamefont {Huang}, \citenamefont {Ioffe}, \citenamefont {Isakov},
  \citenamefont {Jeffrey}, \citenamefont {Jiang}, \citenamefont {Jones},
  \citenamefont {Kafri}, \citenamefont {Kechedzhi}, \citenamefont {Kelly},
  \citenamefont {Kim}, \citenamefont {Klimov}, \citenamefont {Korotkov},
  \citenamefont {Kostritsa}, \citenamefont {Landhuis}, \citenamefont {Laptev},
  \citenamefont {Lindmark}, \citenamefont {Leib}, \citenamefont {Martin},
  \citenamefont {Martinis}, \citenamefont {McClean}, \citenamefont {McEwen},
  \citenamefont {Megrant}, \citenamefont {Mi}, \citenamefont {Mohseni},
  \citenamefont {Mruczkiewicz}, \citenamefont {Mutus}, \citenamefont {Naaman},
  \citenamefont {Neill}, \citenamefont {Neukart}, \citenamefont {Niu},
  \citenamefont {O’Brien}, \citenamefont {O’Gorman}, \citenamefont {Ostby},
  \citenamefont {Petukhov}, \citenamefont {Putterman}, \citenamefont
  {Quintana}, \citenamefont {Roushan}, \citenamefont {Rubin}, \citenamefont
  {Sank}, \citenamefont {Skolik}, \citenamefont {Smelyanskiy}, \citenamefont
  {Strain}, \citenamefont {Streif}, \citenamefont {Szalay}, \citenamefont
  {Vainsencher}, \citenamefont {White}, \citenamefont {Yao}, \citenamefont
  {Yeh}, \citenamefont {Zalcman}, \citenamefont {Zhou}, \citenamefont {Neven},
  \citenamefont {Bacon}, \citenamefont {Lucero}, \citenamefont {Farhi},\ and\
  \citenamefont {Babbush}}]{Harrigan2021QuantumProcessor}%
  \BibitemOpen
  \bibfield  {author} {\bibinfo {author} {\bibfnamefont {M.~P.}\ \bibnamefont
  {Harrigan}}, \bibinfo {author} {\bibfnamefont {K.~J.}\ \bibnamefont {Sung}},
  \bibinfo {author} {\bibfnamefont {M.}~\bibnamefont {Neeley}}, \bibinfo
  {author} {\bibfnamefont {K.~J.}\ \bibnamefont {Satzinger}}, \bibinfo {author}
  {\bibfnamefont {F.}~\bibnamefont {Arute}}, \bibinfo {author} {\bibfnamefont
  {K.}~\bibnamefont {Arya}}, \bibinfo {author} {\bibfnamefont {J.}~\bibnamefont
  {Atalaya}}, \bibinfo {author} {\bibfnamefont {J.~C.}\ \bibnamefont {Bardin}},
  \bibinfo {author} {\bibfnamefont {R.}~\bibnamefont {Barends}}, \bibinfo
  {author} {\bibfnamefont {S.}~\bibnamefont {Boixo}}, \bibinfo {author}
  {\bibfnamefont {M.}~\bibnamefont {Broughton}}, \bibinfo {author}
  {\bibfnamefont {B.~B.}\ \bibnamefont {Buckley}}, \bibinfo {author}
  {\bibfnamefont {D.~A.}\ \bibnamefont {Buell}}, \bibinfo {author}
  {\bibfnamefont {B.}~\bibnamefont {Burkett}}, \bibinfo {author} {\bibfnamefont
  {N.}~\bibnamefont {Bushnell}}, \bibinfo {author} {\bibfnamefont
  {Y.}~\bibnamefont {Chen}}, \bibinfo {author} {\bibfnamefont {Z.}~\bibnamefont
  {Chen}}, \bibinfo {author} {\bibnamefont {{Ben Chiaro}}}, \bibinfo {author}
  {\bibfnamefont {R.}~\bibnamefont {Collins}}, \bibinfo {author} {\bibfnamefont
  {W.}~\bibnamefont {Courtney}}, \bibinfo {author} {\bibfnamefont
  {S.}~\bibnamefont {Demura}}, \bibinfo {author} {\bibfnamefont
  {A.}~\bibnamefont {Dunsworth}}, \bibinfo {author} {\bibfnamefont
  {D.}~\bibnamefont {Eppens}}, \bibinfo {author} {\bibfnamefont
  {A.}~\bibnamefont {Fowler}}, \bibinfo {author} {\bibfnamefont
  {B.}~\bibnamefont {Foxen}}, \bibinfo {author} {\bibfnamefont
  {C.}~\bibnamefont {Gidney}}, \bibinfo {author} {\bibfnamefont
  {M.}~\bibnamefont {Giustina}}, \bibinfo {author} {\bibfnamefont
  {R.}~\bibnamefont {Graff}}, \bibinfo {author} {\bibfnamefont
  {S.}~\bibnamefont {Habegger}}, \bibinfo {author} {\bibfnamefont
  {A.}~\bibnamefont {Ho}}, \bibinfo {author} {\bibfnamefont {S.}~\bibnamefont
  {Hong}}, \bibinfo {author} {\bibfnamefont {T.}~\bibnamefont {Huang}},
  \bibinfo {author} {\bibfnamefont {L.~B.}\ \bibnamefont {Ioffe}}, \bibinfo
  {author} {\bibfnamefont {S.~V.}\ \bibnamefont {Isakov}}, \bibinfo {author}
  {\bibfnamefont {E.}~\bibnamefont {Jeffrey}}, \bibinfo {author} {\bibfnamefont
  {Z.}~\bibnamefont {Jiang}}, \bibinfo {author} {\bibfnamefont
  {C.}~\bibnamefont {Jones}}, \bibinfo {author} {\bibfnamefont
  {D.}~\bibnamefont {Kafri}}, \bibinfo {author} {\bibfnamefont
  {K.}~\bibnamefont {Kechedzhi}}, \bibinfo {author} {\bibfnamefont
  {J.}~\bibnamefont {Kelly}}, \bibinfo {author} {\bibfnamefont
  {S.}~\bibnamefont {Kim}}, \bibinfo {author} {\bibfnamefont {P.~V.}\
  \bibnamefont {Klimov}}, \bibinfo {author} {\bibfnamefont {A.~N.}\
  \bibnamefont {Korotkov}}, \bibinfo {author} {\bibfnamefont {F.}~\bibnamefont
  {Kostritsa}}, \bibinfo {author} {\bibfnamefont {D.}~\bibnamefont {Landhuis}},
  \bibinfo {author} {\bibfnamefont {P.}~\bibnamefont {Laptev}}, \bibinfo
  {author} {\bibfnamefont {M.}~\bibnamefont {Lindmark}}, \bibinfo {author}
  {\bibfnamefont {M.}~\bibnamefont {Leib}}, \bibinfo {author} {\bibfnamefont
  {O.}~\bibnamefont {Martin}}, \bibinfo {author} {\bibfnamefont {J.~M.}\
  \bibnamefont {Martinis}}, \bibinfo {author} {\bibfnamefont {J.~R.}\
  \bibnamefont {McClean}}, \bibinfo {author} {\bibfnamefont {M.}~\bibnamefont
  {McEwen}}, \bibinfo {author} {\bibfnamefont {A.}~\bibnamefont {Megrant}},
  \bibinfo {author} {\bibfnamefont {X.}~\bibnamefont {Mi}}, \bibinfo {author}
  {\bibfnamefont {M.}~\bibnamefont {Mohseni}}, \bibinfo {author} {\bibfnamefont
  {W.}~\bibnamefont {Mruczkiewicz}}, \bibinfo {author} {\bibfnamefont
  {J.}~\bibnamefont {Mutus}}, \bibinfo {author} {\bibfnamefont
  {O.}~\bibnamefont {Naaman}}, \bibinfo {author} {\bibfnamefont
  {C.}~\bibnamefont {Neill}}, \bibinfo {author} {\bibfnamefont
  {F.}~\bibnamefont {Neukart}}, \bibinfo {author} {\bibfnamefont {M.~Y.}\
  \bibnamefont {Niu}}, \bibinfo {author} {\bibfnamefont {T.~E.}\ \bibnamefont
  {O’Brien}}, \bibinfo {author} {\bibfnamefont {B.}~\bibnamefont
  {O’Gorman}}, \bibinfo {author} {\bibfnamefont {E.}~\bibnamefont {Ostby}},
  \bibinfo {author} {\bibfnamefont {A.}~\bibnamefont {Petukhov}}, \bibinfo
  {author} {\bibfnamefont {H.}~\bibnamefont {Putterman}}, \bibinfo {author}
  {\bibfnamefont {C.}~\bibnamefont {Quintana}}, \bibinfo {author}
  {\bibfnamefont {P.}~\bibnamefont {Roushan}}, \bibinfo {author} {\bibfnamefont
  {N.~C.}\ \bibnamefont {Rubin}}, \bibinfo {author} {\bibfnamefont
  {D.}~\bibnamefont {Sank}}, \bibinfo {author} {\bibfnamefont {A.}~\bibnamefont
  {Skolik}}, \bibinfo {author} {\bibfnamefont {V.}~\bibnamefont {Smelyanskiy}},
  \bibinfo {author} {\bibfnamefont {D.}~\bibnamefont {Strain}}, \bibinfo
  {author} {\bibfnamefont {M.}~\bibnamefont {Streif}}, \bibinfo {author}
  {\bibfnamefont {M.}~\bibnamefont {Szalay}}, \bibinfo {author} {\bibfnamefont
  {A.}~\bibnamefont {Vainsencher}}, \bibinfo {author} {\bibfnamefont
  {T.}~\bibnamefont {White}}, \bibinfo {author} {\bibfnamefont {Z.~J.}\
  \bibnamefont {Yao}}, \bibinfo {author} {\bibfnamefont {P.}~\bibnamefont
  {Yeh}}, \bibinfo {author} {\bibfnamefont {A.}~\bibnamefont {Zalcman}},
  \bibinfo {author} {\bibfnamefont {L.}~\bibnamefont {Zhou}}, \bibinfo {author}
  {\bibfnamefont {H.}~\bibnamefont {Neven}}, \bibinfo {author} {\bibfnamefont
  {D.}~\bibnamefont {Bacon}}, \bibinfo {author} {\bibfnamefont
  {E.}~\bibnamefont {Lucero}}, \bibinfo {author} {\bibfnamefont
  {E.}~\bibnamefont {Farhi}},\ and\ \bibinfo {author} {\bibfnamefont
  {R.}~\bibnamefont {Babbush}},\ }\bibfield  {title} {\bibinfo {title}
  {{Quantum approximate optimization of non-planar graph problems on a planar
  superconducting processor}},\ }\bibfield  {journal} {\bibinfo  {journal}
  {Nature Physics}\ }\href {https://doi.org/10.1038/s41567-020-01105-y}
  {10.1038/s41567-020-01105-y} (\bibinfo {year} {2021})\BibitemShut {NoStop}%
\bibitem [{\citenamefont {Kandala}\ \emph {et~al.}(2017)\citenamefont
  {Kandala}, \citenamefont {Mezzacapo}, \citenamefont {Temme}, \citenamefont
  {Takita}, \citenamefont {Brink}, \citenamefont {Chow},\ and\ \citenamefont
  {Gambetta}}]{Kandala2017Hardware-efficientMagnets}%
  \BibitemOpen
  \bibfield  {author} {\bibinfo {author} {\bibfnamefont {A.}~\bibnamefont
  {Kandala}}, \bibinfo {author} {\bibfnamefont {A.}~\bibnamefont {Mezzacapo}},
  \bibinfo {author} {\bibfnamefont {K.}~\bibnamefont {Temme}}, \bibinfo
  {author} {\bibfnamefont {M.}~\bibnamefont {Takita}}, \bibinfo {author}
  {\bibfnamefont {M.}~\bibnamefont {Brink}}, \bibinfo {author} {\bibfnamefont
  {J.~M.}\ \bibnamefont {Chow}},\ and\ \bibinfo {author} {\bibfnamefont
  {J.~M.}\ \bibnamefont {Gambetta}},\ }\bibfield  {title} {\bibinfo {title}
  {{Hardware-efficient variational quantum eigensolver for small molecules and
  quantum magnets}},\ }\href {https://doi.org/10.1038/nature23879} {\bibfield
  {journal} {\bibinfo  {journal} {Nature}\ }\textbf {\bibinfo {volume} {549}},\
  \bibinfo {pages} {242} (\bibinfo {year} {2017})}\BibitemShut {NoStop}%
\bibitem [{\citenamefont {Hashim}\ \emph {et~al.}(2020)\citenamefont {Hashim},
  \citenamefont {Naik}, \citenamefont {Morvan}, \citenamefont {Ville},
  \citenamefont {Mitchell}, \citenamefont {Kreikebaum}, \citenamefont {Davis},
  \citenamefont {Smith}, \citenamefont {Iancu}, \citenamefont {O’Brien},
  \citenamefont {Hincks}, \citenamefont {Wallman}, \citenamefont {Emerson},\
  and\ \citenamefont {Siddiqi}}]{Hashim2020RandomizedProcessor}%
  \BibitemOpen
  \bibfield  {author} {\bibinfo {author} {\bibfnamefont {A.}~\bibnamefont
  {Hashim}}, \bibinfo {author} {\bibfnamefont {R.~K.}\ \bibnamefont {Naik}},
  \bibinfo {author} {\bibfnamefont {A.}~\bibnamefont {Morvan}}, \bibinfo
  {author} {\bibfnamefont {J.~L.}\ \bibnamefont {Ville}}, \bibinfo {author}
  {\bibfnamefont {B.}~\bibnamefont {Mitchell}}, \bibinfo {author}
  {\bibfnamefont {J.~M.}\ \bibnamefont {Kreikebaum}}, \bibinfo {author}
  {\bibfnamefont {M.}~\bibnamefont {Davis}}, \bibinfo {author} {\bibfnamefont
  {E.}~\bibnamefont {Smith}}, \bibinfo {author} {\bibfnamefont
  {C.}~\bibnamefont {Iancu}}, \bibinfo {author} {\bibfnamefont {K.~P.}\
  \bibnamefont {O’Brien}}, \bibinfo {author} {\bibfnamefont {I.}~\bibnamefont
  {Hincks}}, \bibinfo {author} {\bibfnamefont {J.~J.}\ \bibnamefont {Wallman}},
  \bibinfo {author} {\bibfnamefont {J.}~\bibnamefont {Emerson}},\ and\ \bibinfo
  {author} {\bibfnamefont {I.}~\bibnamefont {Siddiqi}},\ }\href@noop {}
  {\bibinfo {title} {{Randomized compiling for scalable quantum computing on a
  noisy superconducting quantum processor}}} (\bibinfo {year}
  {2020})\BibitemShut {NoStop}%
\bibitem [{\citenamefont {Song}\ \emph {et~al.}(2017)\citenamefont {Song},
  \citenamefont {Xu}, \citenamefont {Liu}, \citenamefont {Yang}, \citenamefont
  {Zheng}, \citenamefont {Deng}, \citenamefont {Xie}, \citenamefont {Huang},
  \citenamefont {Guo}, \citenamefont {Zhang}, \citenamefont {Zhang},
  \citenamefont {Xu}, \citenamefont {Zheng}, \citenamefont {Zhu}, \citenamefont
  {Wang}, \citenamefont {Chen}, \citenamefont {Lu}, \citenamefont {Han},\ and\
  \citenamefont {Pan}}]{Song201710-QubitCircuit}%
  \BibitemOpen
  \bibfield  {author} {\bibinfo {author} {\bibfnamefont {C.}~\bibnamefont
  {Song}}, \bibinfo {author} {\bibfnamefont {K.}~\bibnamefont {Xu}}, \bibinfo
  {author} {\bibfnamefont {W.}~\bibnamefont {Liu}}, \bibinfo {author}
  {\bibfnamefont {C.~P.}\ \bibnamefont {Yang}}, \bibinfo {author}
  {\bibfnamefont {S.~B.}\ \bibnamefont {Zheng}}, \bibinfo {author}
  {\bibfnamefont {H.}~\bibnamefont {Deng}}, \bibinfo {author} {\bibfnamefont
  {Q.}~\bibnamefont {Xie}}, \bibinfo {author} {\bibfnamefont {K.}~\bibnamefont
  {Huang}}, \bibinfo {author} {\bibfnamefont {Q.}~\bibnamefont {Guo}}, \bibinfo
  {author} {\bibfnamefont {L.}~\bibnamefont {Zhang}}, \bibinfo {author}
  {\bibfnamefont {P.}~\bibnamefont {Zhang}}, \bibinfo {author} {\bibfnamefont
  {D.}~\bibnamefont {Xu}}, \bibinfo {author} {\bibfnamefont {D.}~\bibnamefont
  {Zheng}}, \bibinfo {author} {\bibfnamefont {X.}~\bibnamefont {Zhu}}, \bibinfo
  {author} {\bibfnamefont {H.}~\bibnamefont {Wang}}, \bibinfo {author}
  {\bibfnamefont {Y.~A.}\ \bibnamefont {Chen}}, \bibinfo {author}
  {\bibfnamefont {C.~Y.}\ \bibnamefont {Lu}}, \bibinfo {author} {\bibfnamefont
  {S.}~\bibnamefont {Han}},\ and\ \bibinfo {author} {\bibfnamefont {J.~W.}\
  \bibnamefont {Pan}},\ }\bibfield  {title} {\bibinfo {title} {{10-Qubit
  Entanglement and Parallel Logic Operations with a Superconducting Circuit}},\
  }\bibfield  {journal} {\bibinfo  {journal} {Physical Review Letters}\
  }\textbf {\bibinfo {volume} {119}},\ \href
  {https://doi.org/10.1103/PhysRevLett.119.180511}
  {10.1103/PhysRevLett.119.180511} (\bibinfo {year} {2017})\BibitemShut
  {NoStop}%
\bibitem [{\citenamefont {Jurcevic}\ \emph {et~al.}(2021)\citenamefont
  {Jurcevic}, \citenamefont {Javadi-Abhari}, \citenamefont {Bishop},
  \citenamefont {Lauer}, \citenamefont {Borgorin}, \citenamefont {Brink},
  \citenamefont {Capelluto}, \citenamefont {Gunluk}, \citenamefont {Itoko},
  \citenamefont {Kanazawa}, \citenamefont {Kandala}, \citenamefont {Keefe},
  \citenamefont {Krsulich}, \citenamefont {Landers}, \citenamefont
  {Lewandowski}, \citenamefont {McClure}, \citenamefont {Nannicini},
  \citenamefont {Narasgond}, \citenamefont {Nayfeh}, \citenamefont {Pritchett},
  \citenamefont {Rothwell}, \citenamefont {Srinivasan}, \citenamefont
  {Sundaresan}, \citenamefont {Wang}, \citenamefont {Wei}, \citenamefont
  {Wood}, \citenamefont {Yau}, \citenamefont {Zhang}, \citenamefont {Dial},
  \citenamefont {Chow},\ and\ \citenamefont
  {Gambetta}}]{Jurcevic2021DemonstrationSystem}%
  \BibitemOpen
  \bibfield  {author} {\bibinfo {author} {\bibfnamefont {P.}~\bibnamefont
  {Jurcevic}}, \bibinfo {author} {\bibfnamefont {A.}~\bibnamefont
  {Javadi-Abhari}}, \bibinfo {author} {\bibfnamefont {L.~S.}\ \bibnamefont
  {Bishop}}, \bibinfo {author} {\bibfnamefont {I.}~\bibnamefont {Lauer}},
  \bibinfo {author} {\bibfnamefont {D.}~\bibnamefont {Borgorin}}, \bibinfo
  {author} {\bibfnamefont {M.}~\bibnamefont {Brink}}, \bibinfo {author}
  {\bibfnamefont {L.}~\bibnamefont {Capelluto}}, \bibinfo {author}
  {\bibfnamefont {O.}~\bibnamefont {Gunluk}}, \bibinfo {author} {\bibfnamefont
  {T.}~\bibnamefont {Itoko}}, \bibinfo {author} {\bibfnamefont
  {N.}~\bibnamefont {Kanazawa}}, \bibinfo {author} {\bibfnamefont
  {A.}~\bibnamefont {Kandala}}, \bibinfo {author} {\bibfnamefont
  {G.}~\bibnamefont {Keefe}}, \bibinfo {author} {\bibfnamefont
  {K.}~\bibnamefont {Krsulich}}, \bibinfo {author} {\bibfnamefont
  {W.}~\bibnamefont {Landers}}, \bibinfo {author} {\bibfnamefont {E.~P.}\
  \bibnamefont {Lewandowski}}, \bibinfo {author} {\bibfnamefont {D.~T.}\
  \bibnamefont {McClure}}, \bibinfo {author} {\bibfnamefont {G.}~\bibnamefont
  {Nannicini}}, \bibinfo {author} {\bibfnamefont {A.}~\bibnamefont
  {Narasgond}}, \bibinfo {author} {\bibfnamefont {H.~M.}\ \bibnamefont
  {Nayfeh}}, \bibinfo {author} {\bibfnamefont {E.}~\bibnamefont {Pritchett}},
  \bibinfo {author} {\bibfnamefont {M.~B.}\ \bibnamefont {Rothwell}}, \bibinfo
  {author} {\bibfnamefont {S.}~\bibnamefont {Srinivasan}}, \bibinfo {author}
  {\bibfnamefont {N.}~\bibnamefont {Sundaresan}}, \bibinfo {author}
  {\bibfnamefont {C.}~\bibnamefont {Wang}}, \bibinfo {author} {\bibfnamefont
  {K.~X.}\ \bibnamefont {Wei}}, \bibinfo {author} {\bibfnamefont {C.~J.}\
  \bibnamefont {Wood}}, \bibinfo {author} {\bibfnamefont {J.-b.}\ \bibnamefont
  {Yau}}, \bibinfo {author} {\bibfnamefont {E.}~\bibnamefont {Zhang}}, \bibinfo
  {author} {\bibfnamefont {O.~E.}\ \bibnamefont {Dial}}, \bibinfo {author}
  {\bibfnamefont {J.}~\bibnamefont {Chow}},\ and\ \bibinfo {author}
  {\bibfnamefont {J.}~\bibnamefont {Gambetta}},\ }\bibfield  {title} {\bibinfo
  {title} {{Demonstration of quantum volume 64 on a superconducting quantum
  computing system}},\ }\bibfield  {journal} {\bibinfo  {journal} {Quantum
  Science and Technology}\ }\href {https://doi.org/10.1088/2058-9565/abe519}
  {10.1088/2058-9565/abe519} (\bibinfo {year} {2021})\BibitemShut {NoStop}%
\bibitem [{\citenamefont {Barends}\ \emph {et~al.}(2015)\citenamefont
  {Barends}, \citenamefont {Lamata}, \citenamefont {Kelly}, \citenamefont
  {Garc{\'{i}}a-{\'{A}}lvarez}, \citenamefont {Fowler}, \citenamefont
  {Megrant}, \citenamefont {Jeffrey}, \citenamefont {White}, \citenamefont
  {Sank}, \citenamefont {Mutus}, \citenamefont {Campbell}, \citenamefont
  {Chen}, \citenamefont {Chen}, \citenamefont {Chiaro}, \citenamefont
  {Dunsworth}, \citenamefont {Hoi}, \citenamefont {Neill}, \citenamefont
  {O'Malley}, \citenamefont {Quintana}, \citenamefont {Roushan}, \citenamefont
  {Vainsencher}, \citenamefont {Wenner}, \citenamefont {Solano},\ and\
  \citenamefont {Martinis}}]{Barends2015DigitalCircuit}%
  \BibitemOpen
  \bibfield  {author} {\bibinfo {author} {\bibfnamefont {R.}~\bibnamefont
  {Barends}}, \bibinfo {author} {\bibfnamefont {L.}~\bibnamefont {Lamata}},
  \bibinfo {author} {\bibfnamefont {J.}~\bibnamefont {Kelly}}, \bibinfo
  {author} {\bibfnamefont {L.}~\bibnamefont {Garc{\'{i}}a-{\'{A}}lvarez}},
  \bibinfo {author} {\bibfnamefont {A.~G.}\ \bibnamefont {Fowler}}, \bibinfo
  {author} {\bibfnamefont {A.}~\bibnamefont {Megrant}}, \bibinfo {author}
  {\bibfnamefont {E.}~\bibnamefont {Jeffrey}}, \bibinfo {author} {\bibfnamefont
  {T.~C.}\ \bibnamefont {White}}, \bibinfo {author} {\bibfnamefont
  {D.}~\bibnamefont {Sank}}, \bibinfo {author} {\bibfnamefont {J.~Y.}\
  \bibnamefont {Mutus}}, \bibinfo {author} {\bibfnamefont {B.}~\bibnamefont
  {Campbell}}, \bibinfo {author} {\bibfnamefont {Y.}~\bibnamefont {Chen}},
  \bibinfo {author} {\bibfnamefont {Z.}~\bibnamefont {Chen}}, \bibinfo {author}
  {\bibfnamefont {B.}~\bibnamefont {Chiaro}}, \bibinfo {author} {\bibfnamefont
  {A.}~\bibnamefont {Dunsworth}}, \bibinfo {author} {\bibfnamefont {I.~C.}\
  \bibnamefont {Hoi}}, \bibinfo {author} {\bibfnamefont {C.}~\bibnamefont
  {Neill}}, \bibinfo {author} {\bibfnamefont {P.~J.}\ \bibnamefont {O'Malley}},
  \bibinfo {author} {\bibfnamefont {C.}~\bibnamefont {Quintana}}, \bibinfo
  {author} {\bibfnamefont {P.}~\bibnamefont {Roushan}}, \bibinfo {author}
  {\bibfnamefont {A.}~\bibnamefont {Vainsencher}}, \bibinfo {author}
  {\bibfnamefont {J.}~\bibnamefont {Wenner}}, \bibinfo {author} {\bibfnamefont
  {E.}~\bibnamefont {Solano}},\ and\ \bibinfo {author} {\bibfnamefont {J.~M.}\
  \bibnamefont {Martinis}},\ }\bibfield  {title} {\bibinfo {title} {{Digital
  quantum simulation of fermionic models with a superconducting circuit}},\
  }\bibfield  {journal} {\bibinfo  {journal} {Nature Communications}\ }\textbf
  {\bibinfo {volume} {6}},\ \href {https://doi.org/10.1038/ncomms8654}
  {10.1038/ncomms8654} (\bibinfo {year} {2015})\BibitemShut {NoStop}%
\bibitem [{\citenamefont {Salath{\'{e}}}\ \emph {et~al.}(2015)\citenamefont
  {Salath{\'{e}}}, \citenamefont {Mondal}, \citenamefont {Oppliger},
  \citenamefont {Heinsoo}, \citenamefont {Kurpiers}, \citenamefont
  {Poto{\v{c}}nik}, \citenamefont {Mezzacapo}, \citenamefont {Las~Heras},
  \citenamefont {Lamata}, \citenamefont {Solano}, \citenamefont {Filipp},\ and\
  \citenamefont {Wallraff}}]{Salathe2015DigitalElectrodynamics}%
  \BibitemOpen
  \bibfield  {author} {\bibinfo {author} {\bibfnamefont {Y.}~\bibnamefont
  {Salath{\'{e}}}}, \bibinfo {author} {\bibfnamefont {M.}~\bibnamefont
  {Mondal}}, \bibinfo {author} {\bibfnamefont {M.}~\bibnamefont {Oppliger}},
  \bibinfo {author} {\bibfnamefont {J.}~\bibnamefont {Heinsoo}}, \bibinfo
  {author} {\bibfnamefont {P.}~\bibnamefont {Kurpiers}}, \bibinfo {author}
  {\bibfnamefont {A.}~\bibnamefont {Poto{\v{c}}nik}}, \bibinfo {author}
  {\bibfnamefont {A.}~\bibnamefont {Mezzacapo}}, \bibinfo {author}
  {\bibfnamefont {U.}~\bibnamefont {Las~Heras}}, \bibinfo {author}
  {\bibfnamefont {L.}~\bibnamefont {Lamata}}, \bibinfo {author} {\bibfnamefont
  {E.}~\bibnamefont {Solano}}, \bibinfo {author} {\bibfnamefont
  {S.}~\bibnamefont {Filipp}},\ and\ \bibinfo {author} {\bibfnamefont
  {A.}~\bibnamefont {Wallraff}},\ }\bibfield  {title} {\bibinfo {title}
  {{Digital quantum simulation of spin models with circuit quantum
  electrodynamics}},\ }\bibfield  {journal} {\bibinfo  {journal} {Physical
  Review X}\ }\textbf {\bibinfo {volume} {5}},\ \href
  {https://doi.org/10.1103/PhysRevX.5.021027} {10.1103/PhysRevX.5.021027}
  (\bibinfo {year} {2015})\BibitemShut {NoStop}%
\bibitem [{\citenamefont {Ofek}\ \emph {et~al.}(2016)\citenamefont {Ofek},
  \citenamefont {Petrenko}, \citenamefont {Heeres}, \citenamefont {Reinhold},
  \citenamefont {Leghtas}, \citenamefont {Vlastakis}, \citenamefont {Liu},
  \citenamefont {Frunzio}, \citenamefont {Girvin}, \citenamefont {Jiang},
  \citenamefont {Mirrahimi}, \citenamefont {Devoret},\ and\ \citenamefont
  {Schoelkopf}}]{Ofek2016ExtendingCircuits}%
  \BibitemOpen
  \bibfield  {author} {\bibinfo {author} {\bibfnamefont {N.}~\bibnamefont
  {Ofek}}, \bibinfo {author} {\bibfnamefont {A.}~\bibnamefont {Petrenko}},
  \bibinfo {author} {\bibfnamefont {R.}~\bibnamefont {Heeres}}, \bibinfo
  {author} {\bibfnamefont {P.}~\bibnamefont {Reinhold}}, \bibinfo {author}
  {\bibfnamefont {Z.}~\bibnamefont {Leghtas}}, \bibinfo {author} {\bibfnamefont
  {B.}~\bibnamefont {Vlastakis}}, \bibinfo {author} {\bibfnamefont
  {Y.}~\bibnamefont {Liu}}, \bibinfo {author} {\bibfnamefont {L.}~\bibnamefont
  {Frunzio}}, \bibinfo {author} {\bibfnamefont {S.~M.}\ \bibnamefont {Girvin}},
  \bibinfo {author} {\bibfnamefont {L.}~\bibnamefont {Jiang}}, \bibinfo
  {author} {\bibfnamefont {M.}~\bibnamefont {Mirrahimi}}, \bibinfo {author}
  {\bibfnamefont {M.~H.}\ \bibnamefont {Devoret}},\ and\ \bibinfo {author}
  {\bibfnamefont {R.~J.}\ \bibnamefont {Schoelkopf}},\ }\bibfield  {title}
  {\bibinfo {title} {{Extending the lifetime of a quantum bit with error
  correction in superconducting circuits}},\ }\href
  {https://doi.org/10.1038/nature18949} {\bibfield  {journal} {\bibinfo
  {journal} {Nature}\ }\textbf {\bibinfo {volume} {536}},\ \bibinfo {pages}
  {441} (\bibinfo {year} {2016})}\BibitemShut {NoStop}%
\bibitem [{\citenamefont {Campagne-Ibarcq}\ \emph {et~al.}(2020)\citenamefont
  {Campagne-Ibarcq}, \citenamefont {Eickbusch}, \citenamefont {Touzard},
  \citenamefont {Zalys-Geller}, \citenamefont {Frattini}, \citenamefont
  {Sivak}, \citenamefont {Reinhold}, \citenamefont {Puri}, \citenamefont
  {Shankar}, \citenamefont {Schoelkopf}, \citenamefont {Frunzio}, \citenamefont
  {Mirrahimi},\ and\ \citenamefont
  {Devoret}}]{Campagne-Ibarcq2020QuantumOscillator}%
  \BibitemOpen
  \bibfield  {author} {\bibinfo {author} {\bibfnamefont {P.}~\bibnamefont
  {Campagne-Ibarcq}}, \bibinfo {author} {\bibfnamefont {A.}~\bibnamefont
  {Eickbusch}}, \bibinfo {author} {\bibfnamefont {S.}~\bibnamefont {Touzard}},
  \bibinfo {author} {\bibfnamefont {E.}~\bibnamefont {Zalys-Geller}}, \bibinfo
  {author} {\bibfnamefont {N.~E.}\ \bibnamefont {Frattini}}, \bibinfo {author}
  {\bibfnamefont {V.~V.}\ \bibnamefont {Sivak}}, \bibinfo {author}
  {\bibfnamefont {P.}~\bibnamefont {Reinhold}}, \bibinfo {author}
  {\bibfnamefont {S.}~\bibnamefont {Puri}}, \bibinfo {author} {\bibfnamefont
  {S.}~\bibnamefont {Shankar}}, \bibinfo {author} {\bibfnamefont {R.~J.}\
  \bibnamefont {Schoelkopf}}, \bibinfo {author} {\bibfnamefont
  {L.}~\bibnamefont {Frunzio}}, \bibinfo {author} {\bibfnamefont
  {M.}~\bibnamefont {Mirrahimi}},\ and\ \bibinfo {author} {\bibfnamefont
  {M.~H.}\ \bibnamefont {Devoret}},\ }\bibfield  {title} {\bibinfo {title}
  {{Quantum error correction of a qubit encoded in grid states of an
  oscillator}},\ }\href {https://doi.org/10.1038/s41586-020-2603-3} {\bibfield
  {journal} {\bibinfo  {journal} {Nature}\ }\textbf {\bibinfo {volume} {584}},\
  \bibinfo {pages} {368} (\bibinfo {year} {2020})}\BibitemShut {NoStop}%
\bibitem [{\citenamefont {Andersen}\ \emph {et~al.}(2020)\citenamefont
  {Andersen}, \citenamefont {Remm}, \citenamefont {Lazar}, \citenamefont
  {Krinner}, \citenamefont {Lacroix}, \citenamefont {Norris}, \citenamefont
  {Gabureac}, \citenamefont {Eichler},\ and\ \citenamefont
  {Wallraff}}]{Andersen2020RepeatedCode}%
  \BibitemOpen
  \bibfield  {author} {\bibinfo {author} {\bibfnamefont {C.~K.}\ \bibnamefont
  {Andersen}}, \bibinfo {author} {\bibfnamefont {A.}~\bibnamefont {Remm}},
  \bibinfo {author} {\bibfnamefont {S.}~\bibnamefont {Lazar}}, \bibinfo
  {author} {\bibfnamefont {S.}~\bibnamefont {Krinner}}, \bibinfo {author}
  {\bibfnamefont {N.}~\bibnamefont {Lacroix}}, \bibinfo {author} {\bibfnamefont
  {G.~J.}\ \bibnamefont {Norris}}, \bibinfo {author} {\bibfnamefont
  {M.}~\bibnamefont {Gabureac}}, \bibinfo {author} {\bibfnamefont
  {C.}~\bibnamefont {Eichler}},\ and\ \bibinfo {author} {\bibfnamefont
  {A.}~\bibnamefont {Wallraff}},\ }\bibfield  {title} {\bibinfo {title}
  {{Repeated quantum error detection in a surface code}},\ }\href
  {https://doi.org/10.1038/s41567-020-0920-y} {\bibfield  {journal} {\bibinfo
  {journal} {Nature Physics}\ }\textbf {\bibinfo {volume} {16}},\ \bibinfo
  {pages} {875} (\bibinfo {year} {2020})}\BibitemShut {NoStop}%
\bibitem [{\citenamefont {Marques}\ \emph {et~al.}(2021)\citenamefont
  {Marques}, \citenamefont {Varbanov}, \citenamefont {Moreira}, \citenamefont
  {Ali}, \citenamefont {Muthusubramanian}, \citenamefont {Zachariadis},
  \citenamefont {Battistel}, \citenamefont {Beekman}, \citenamefont {Haider},
  \citenamefont {Vlothuizen}, \citenamefont {Bruno}, \citenamefont {Terhal},\
  and\ \citenamefont {DiCarlo}}]{Marques2021Logical-qubitCode}%
  \BibitemOpen
  \bibfield  {author} {\bibinfo {author} {\bibfnamefont {J.~F.}\ \bibnamefont
  {Marques}}, \bibinfo {author} {\bibfnamefont {B.~M.}\ \bibnamefont
  {Varbanov}}, \bibinfo {author} {\bibfnamefont {M.~S.}\ \bibnamefont
  {Moreira}}, \bibinfo {author} {\bibfnamefont {H.}~\bibnamefont {Ali}},
  \bibinfo {author} {\bibfnamefont {N.}~\bibnamefont {Muthusubramanian}},
  \bibinfo {author} {\bibfnamefont {C.}~\bibnamefont {Zachariadis}}, \bibinfo
  {author} {\bibfnamefont {F.}~\bibnamefont {Battistel}}, \bibinfo {author}
  {\bibfnamefont {M.}~\bibnamefont {Beekman}}, \bibinfo {author} {\bibfnamefont
  {N.}~\bibnamefont {Haider}}, \bibinfo {author} {\bibfnamefont
  {W.}~\bibnamefont {Vlothuizen}}, \bibinfo {author} {\bibfnamefont
  {A.}~\bibnamefont {Bruno}}, \bibinfo {author} {\bibfnamefont {B.~M.}\
  \bibnamefont {Terhal}},\ and\ \bibinfo {author} {\bibfnamefont
  {L.}~\bibnamefont {DiCarlo}},\ }\bibfield  {title} {\bibinfo {title}
  {{Logical-qubit operations in an error-detecting surface code}},\ }\href
  {http://arxiv.org/abs/2102.13071} {\bibfield  {journal} {\bibinfo  {journal}
  {arXiv:2102.13071}\ } (\bibinfo {year} {2021})}\BibitemShut {NoStop}%
\bibitem [{\citenamefont {Manucharyan}\ \emph {et~al.}(2009)\citenamefont
  {Manucharyan}, \citenamefont {Koch}, \citenamefont {Glazman},\ and\
  \citenamefont {Devoret}}]{Manucharyan2009Fluxonium:Offsets}%
  \BibitemOpen
  \bibfield  {author} {\bibinfo {author} {\bibfnamefont {V.~E.}\ \bibnamefont
  {Manucharyan}}, \bibinfo {author} {\bibfnamefont {J.}~\bibnamefont {Koch}},
  \bibinfo {author} {\bibfnamefont {L.~I.}\ \bibnamefont {Glazman}},\ and\
  \bibinfo {author} {\bibfnamefont {M.~H.}\ \bibnamefont {Devoret}},\
  }\bibfield  {title} {\bibinfo {title} {{Fluxonium: Single cooper-pair circuit
  free of charge offsets}},\ }\bibfield  {journal} {\bibinfo  {journal}
  {Science}\ }\textbf {\bibinfo {volume} {326}},\ \href
  {https://doi.org/10.1126/science.1175552} {10.1126/science.1175552} (\bibinfo
  {year} {2009})\BibitemShut {NoStop}%
\bibitem [{\citenamefont {Nguyen}\ \emph {et~al.}(2019)\citenamefont {Nguyen},
  \citenamefont {Lin}, \citenamefont {Somoroff}, \citenamefont {Mencia},
  \citenamefont {Grabon},\ and\ \citenamefont
  {Manucharyan}}]{Nguyen2019High-CoherenceQubit}%
  \BibitemOpen
  \bibfield  {author} {\bibinfo {author} {\bibfnamefont {L.~B.}\ \bibnamefont
  {Nguyen}}, \bibinfo {author} {\bibfnamefont {Y.~H.}\ \bibnamefont {Lin}},
  \bibinfo {author} {\bibfnamefont {A.}~\bibnamefont {Somoroff}}, \bibinfo
  {author} {\bibfnamefont {R.}~\bibnamefont {Mencia}}, \bibinfo {author}
  {\bibfnamefont {N.}~\bibnamefont {Grabon}},\ and\ \bibinfo {author}
  {\bibfnamefont {V.~E.}\ \bibnamefont {Manucharyan}},\ }\bibfield  {title}
  {\bibinfo {title} {{High-Coherence Fluxonium Qubit}},\ }\bibfield  {journal}
  {\bibinfo  {journal} {Physical Review X}\ }\textbf {\bibinfo {volume} {9}},\
  \href {https://doi.org/10.1103/PhysRevX.9.041041} {10.1103/PhysRevX.9.041041}
  (\bibinfo {year} {2019})\BibitemShut {NoStop}%
\bibitem [{\citenamefont {Lin}\ \emph {et~al.}(2018)\citenamefont {Lin},
  \citenamefont {Nguyen}, \citenamefont {Grabon}, \citenamefont {San~Miguel},
  \citenamefont {Pankratova},\ and\ \citenamefont
  {Manucharyan}}]{Lin2018DemonstrationDecay}%
  \BibitemOpen
  \bibfield  {author} {\bibinfo {author} {\bibfnamefont {Y.-H.}\ \bibnamefont
  {Lin}}, \bibinfo {author} {\bibfnamefont {L.~B.}\ \bibnamefont {Nguyen}},
  \bibinfo {author} {\bibfnamefont {N.}~\bibnamefont {Grabon}}, \bibinfo
  {author} {\bibfnamefont {J.}~\bibnamefont {San~Miguel}}, \bibinfo {author}
  {\bibfnamefont {N.}~\bibnamefont {Pankratova}},\ and\ \bibinfo {author}
  {\bibfnamefont {V.~E.}\ \bibnamefont {Manucharyan}},\ }\bibfield  {title}
  {\bibinfo {title} {{Demonstration of Protection of a Superconducting Qubit
  from Energy Decay}},\ }\href {https://doi.org/10.1103/PhysRevLett.120.150503}
  {\bibfield  {journal} {\bibinfo  {journal} {Physical Review Letters}\
  }\textbf {\bibinfo {volume} {120}},\ \bibinfo {pages} {150503} (\bibinfo
  {year} {2018})}\BibitemShut {NoStop}%
\bibitem [{\citenamefont {Blais}\ \emph {et~al.}(2004)\citenamefont {Blais},
  \citenamefont {Huang}, \citenamefont {Wallraff}, \citenamefont {Girvin},\
  and\ \citenamefont {Schoelkopf}}]{Blais2004CavityComputation}%
  \BibitemOpen
  \bibfield  {author} {\bibinfo {author} {\bibfnamefont {A.}~\bibnamefont
  {Blais}}, \bibinfo {author} {\bibfnamefont {R.-S.}\ \bibnamefont {Huang}},
  \bibinfo {author} {\bibfnamefont {A.}~\bibnamefont {Wallraff}}, \bibinfo
  {author} {\bibfnamefont {S.~M.}\ \bibnamefont {Girvin}},\ and\ \bibinfo
  {author} {\bibfnamefont {R.~J.}\ \bibnamefont {Schoelkopf}},\ }\bibfield
  {title} {\bibinfo {title} {{Cavity quantum electrodynamics for
  superconducting electrical circuits: An architecture for quantum
  computation}},\ }\href {https://doi.org/10.1103/PhysRevA.69.062320}
  {\bibfield  {journal} {\bibinfo  {journal} {Physical Review A}\ }\textbf
  {\bibinfo {volume} {69}},\ \bibinfo {pages} {062320} (\bibinfo {year}
  {2004})}\BibitemShut {NoStop}%
\bibitem [{\citenamefont {Wallraff}\ \emph {et~al.}(2004)\citenamefont
  {Wallraff}, \citenamefont {Schuster}, \citenamefont {Blais}, \citenamefont
  {Frunzio}, \citenamefont {Huang}, \citenamefont {Majer}, \citenamefont
  {Kumar}, \citenamefont {Girvin},\ and\ \citenamefont
  {Schoelkopf}}]{Wallraff2004StrongElectrodynamics}%
  \BibitemOpen
  \bibfield  {author} {\bibinfo {author} {\bibfnamefont {A.}~\bibnamefont
  {Wallraff}}, \bibinfo {author} {\bibfnamefont {D.~I.}\ \bibnamefont
  {Schuster}}, \bibinfo {author} {\bibfnamefont {A.}~\bibnamefont {Blais}},
  \bibinfo {author} {\bibfnamefont {L.}~\bibnamefont {Frunzio}}, \bibinfo
  {author} {\bibfnamefont {R.~S.}\ \bibnamefont {Huang}}, \bibinfo {author}
  {\bibfnamefont {J.}~\bibnamefont {Majer}}, \bibinfo {author} {\bibfnamefont
  {S.}~\bibnamefont {Kumar}}, \bibinfo {author} {\bibfnamefont {S.~M.}\
  \bibnamefont {Girvin}},\ and\ \bibinfo {author} {\bibfnamefont {R.~J.}\
  \bibnamefont {Schoelkopf}},\ }\bibfield  {title} {\bibinfo {title} {{Strong
  coupling of a single photon to a superconducting qubit using circuit quantum
  electrodynamics}},\ }\bibfield  {journal} {\bibinfo  {journal} {Nature}\
  }\textbf {\bibinfo {volume} {431}},\ \href
  {https://doi.org/10.1038/nature02851} {10.1038/nature02851} (\bibinfo {year}
  {2004})\BibitemShut {NoStop}%
\bibitem [{\citenamefont {Blais}\ \emph {et~al.}(2020)\citenamefont {Blais},
  \citenamefont {Grimsmo}, \citenamefont {Girvin},\ and\ \citenamefont
  {Wallraff}}]{Blais2020CircuitElectrodynamics}%
  \BibitemOpen
  \bibfield  {author} {\bibinfo {author} {\bibfnamefont {A.}~\bibnamefont
  {Blais}}, \bibinfo {author} {\bibfnamefont {A.~L.}\ \bibnamefont {Grimsmo}},
  \bibinfo {author} {\bibfnamefont {S.~M.}\ \bibnamefont {Girvin}},\ and\
  \bibinfo {author} {\bibfnamefont {A.}~\bibnamefont {Wallraff}},\ }\href
  {https://doi.org/10.1142/9789811201400{\_}0006} {\bibinfo {title} {{Circuit
  quantum electrodynamics}}} (\bibinfo {year} {2020})\BibitemShut {NoStop}%
\bibitem [{\citenamefont {Zhang}\ \emph {et~al.}(2020)\citenamefont {Zhang},
  \citenamefont {Chakram}, \citenamefont {Roy}, \citenamefont {Earnest},
  \citenamefont {Lu}, \citenamefont {Huang}, \citenamefont {Weiss},
  \citenamefont {Koch},\ and\ \citenamefont
  {Schuster}}]{Zhang2020UniversalQubit}%
  \BibitemOpen
  \bibfield  {author} {\bibinfo {author} {\bibfnamefont {H.}~\bibnamefont
  {Zhang}}, \bibinfo {author} {\bibfnamefont {S.}~\bibnamefont {Chakram}},
  \bibinfo {author} {\bibfnamefont {T.}~\bibnamefont {Roy}}, \bibinfo {author}
  {\bibfnamefont {N.}~\bibnamefont {Earnest}}, \bibinfo {author} {\bibfnamefont
  {Y.}~\bibnamefont {Lu}}, \bibinfo {author} {\bibfnamefont {Z.}~\bibnamefont
  {Huang}}, \bibinfo {author} {\bibfnamefont {D.}~\bibnamefont {Weiss}},
  \bibinfo {author} {\bibfnamefont {J.}~\bibnamefont {Koch}},\ and\ \bibinfo
  {author} {\bibfnamefont {D.~I.}\ \bibnamefont {Schuster}},\ }\href
  {https://doi.org/10.1103/physrevx.11.011010} {\bibinfo {title} {{Universal
  fast flux control of a coherent, low-frequency qubit}}} (\bibinfo {year}
  {2020})\BibitemShut {NoStop}%
\bibitem [{\citenamefont {Wang}\ \emph {et~al.}(2015)\citenamefont {Wang},
  \citenamefont {Axline}, \citenamefont {Gao}, \citenamefont {Brecht},
  \citenamefont {Chu}, \citenamefont {Frunzio}, \citenamefont {Devoret},\ and\
  \citenamefont {Schoelkopf}}]{Wang2015SurfaceQubits}%
  \BibitemOpen
  \bibfield  {author} {\bibinfo {author} {\bibfnamefont {C.}~\bibnamefont
  {Wang}}, \bibinfo {author} {\bibfnamefont {C.}~\bibnamefont {Axline}},
  \bibinfo {author} {\bibfnamefont {Y.~Y.}\ \bibnamefont {Gao}}, \bibinfo
  {author} {\bibfnamefont {T.}~\bibnamefont {Brecht}}, \bibinfo {author}
  {\bibfnamefont {Y.}~\bibnamefont {Chu}}, \bibinfo {author} {\bibfnamefont
  {L.}~\bibnamefont {Frunzio}}, \bibinfo {author} {\bibfnamefont {M.~H.}\
  \bibnamefont {Devoret}},\ and\ \bibinfo {author} {\bibfnamefont {R.~J.}\
  \bibnamefont {Schoelkopf}},\ }\bibfield  {title} {\bibinfo {title} {{Surface
  participation and dielectric loss in superconducting qubits}},\ }\bibfield
  {journal} {\bibinfo  {journal} {Applied Physics Letters}\ }\textbf {\bibinfo
  {volume} {107}},\ \href {https://doi.org/10.1063/1.4934486}
  {10.1063/1.4934486} (\bibinfo {year} {2015})\BibitemShut {NoStop}%
\bibitem [{\citenamefont {Nesterov}\ \emph {et~al.}(2018)\citenamefont
  {Nesterov}, \citenamefont {Pechenezhskiy}, \citenamefont {Wang},
  \citenamefont {Manucharyan},\ and\ \citenamefont
  {Vavilov}}]{Nesterov2018Microwave-activatedQubits}%
  \BibitemOpen
  \bibfield  {author} {\bibinfo {author} {\bibfnamefont {K.~N.}\ \bibnamefont
  {Nesterov}}, \bibinfo {author} {\bibfnamefont {I.~V.}\ \bibnamefont
  {Pechenezhskiy}}, \bibinfo {author} {\bibfnamefont {C.}~\bibnamefont {Wang}},
  \bibinfo {author} {\bibfnamefont {V.~E.}\ \bibnamefont {Manucharyan}},\ and\
  \bibinfo {author} {\bibfnamefont {M.~G.}\ \bibnamefont {Vavilov}},\
  }\bibfield  {title} {\bibinfo {title} {{Microwave-activated controlled- Z
  gate for fixed-frequency fluxonium qubits}},\ }\bibfield  {journal} {\bibinfo
   {journal} {Physical Review A}\ }\textbf {\bibinfo {volume} {98}},\ \href
  {https://doi.org/10.1103/PhysRevA.98.030301} {10.1103/PhysRevA.98.030301}
  (\bibinfo {year} {2018})\BibitemShut {NoStop}%
\bibitem [{\citenamefont {Ficheux}\ \emph {et~al.}(2020)\citenamefont
  {Ficheux}, \citenamefont {Nguyen}, \citenamefont {Somoroff}, \citenamefont
  {Xiong}, \citenamefont {Nesterov}, \citenamefont {Vavilov},\ and\
  \citenamefont {Manucharyan}}]{Ficheux2020FastFluxoniums}%
  \BibitemOpen
  \bibfield  {author} {\bibinfo {author} {\bibfnamefont {Q.}~\bibnamefont
  {Ficheux}}, \bibinfo {author} {\bibfnamefont {L.~B.}\ \bibnamefont {Nguyen}},
  \bibinfo {author} {\bibfnamefont {A.}~\bibnamefont {Somoroff}}, \bibinfo
  {author} {\bibfnamefont {H.}~\bibnamefont {Xiong}}, \bibinfo {author}
  {\bibfnamefont {K.~N.}\ \bibnamefont {Nesterov}}, \bibinfo {author}
  {\bibfnamefont {M.~G.}\ \bibnamefont {Vavilov}},\ and\ \bibinfo {author}
  {\bibfnamefont {V.~E.}\ \bibnamefont {Manucharyan}},\ }\bibfield  {title}
  {\bibinfo {title} {{Fast logic with slow qubits: microwave-activated
  controlled-Z gate on low-frequency fluxoniums}},\ }\href
  {http://arxiv.org/abs/2011.02634} {\bibfield  {journal} {\bibinfo  {journal}
  {arXiv:2011.02634}\ } (\bibinfo {year} {2020})}\BibitemShut {NoStop}%
\bibitem [{\citenamefont {Xiong}\ \emph {et~al.}(2021)\citenamefont {Xiong},
  \citenamefont {Ficheux}, \citenamefont {Somoroff}, \citenamefont {Nguyen},
  \citenamefont {Dogan}, \citenamefont {Rosenstock}, \citenamefont {Wang},
  \citenamefont {Nesterov}, \citenamefont {Vavilov},\ and\ \citenamefont
  {Manucharyan}}]{Xiong2021ArbitraryShifts}%
  \BibitemOpen
  \bibfield  {author} {\bibinfo {author} {\bibfnamefont {H.}~\bibnamefont
  {Xiong}}, \bibinfo {author} {\bibfnamefont {Q.}~\bibnamefont {Ficheux}},
  \bibinfo {author} {\bibfnamefont {A.}~\bibnamefont {Somoroff}}, \bibinfo
  {author} {\bibfnamefont {L.~B.}\ \bibnamefont {Nguyen}}, \bibinfo {author}
  {\bibfnamefont {E.}~\bibnamefont {Dogan}}, \bibinfo {author} {\bibfnamefont
  {D.}~\bibnamefont {Rosenstock}}, \bibinfo {author} {\bibfnamefont
  {C.}~\bibnamefont {Wang}}, \bibinfo {author} {\bibfnamefont {K.~N.}\
  \bibnamefont {Nesterov}}, \bibinfo {author} {\bibfnamefont {M.~G.}\
  \bibnamefont {Vavilov}},\ and\ \bibinfo {author} {\bibfnamefont {V.~E.}\
  \bibnamefont {Manucharyan}},\ }\bibfield  {title} {\bibinfo {title}
  {{Arbitrary controlled-phase gate on fluxonium qubits using differential
  ac-Stark shifts}},\ }\href {http://arxiv.org/abs/2103.04491} {\bibfield
  {journal} {\bibinfo  {journal} {arXiv:2103.04491}\ } (\bibinfo {year}
  {2021})}\BibitemShut {NoStop}%
\bibitem [{\citenamefont {Magesan}\ \emph {et~al.}(2011)\citenamefont
  {Magesan}, \citenamefont {Gambetta},\ and\ \citenamefont
  {Emerson}}]{Magesan2011ScalableProcesses}%
  \BibitemOpen
  \bibfield  {author} {\bibinfo {author} {\bibfnamefont {E.}~\bibnamefont
  {Magesan}}, \bibinfo {author} {\bibfnamefont {J.~M.}\ \bibnamefont
  {Gambetta}},\ and\ \bibinfo {author} {\bibfnamefont {J.}~\bibnamefont
  {Emerson}},\ }\bibfield  {title} {\bibinfo {title} {{Scalable and robust
  randomized benchmarking of quantum processes}},\ }\bibfield  {journal}
  {\bibinfo  {journal} {Physical Review Letters}\ }\textbf {\bibinfo {volume}
  {106}},\ \href {https://doi.org/10.1103/PhysRevLett.106.180504}
  {10.1103/PhysRevLett.106.180504} (\bibinfo {year} {2011})\BibitemShut
  {NoStop}%
\bibitem [{\citenamefont {Magesan}\ \emph {et~al.}(2012)\citenamefont
  {Magesan}, \citenamefont {Gambetta}, \citenamefont {Johnson}, \citenamefont
  {Ryan}, \citenamefont {Chow}, \citenamefont {Merkel}, \citenamefont
  {Da~Silva}, \citenamefont {Keefe}, \citenamefont {Rothwell}, \citenamefont
  {Ohki}, \citenamefont {Ketchen},\ and\ \citenamefont
  {Steffen}}]{Magesan2012EfficientBenchmarking}%
  \BibitemOpen
  \bibfield  {author} {\bibinfo {author} {\bibfnamefont {E.}~\bibnamefont
  {Magesan}}, \bibinfo {author} {\bibfnamefont {J.~M.}\ \bibnamefont
  {Gambetta}}, \bibinfo {author} {\bibfnamefont {B.~R.}\ \bibnamefont
  {Johnson}}, \bibinfo {author} {\bibfnamefont {C.~A.}\ \bibnamefont {Ryan}},
  \bibinfo {author} {\bibfnamefont {J.~M.}\ \bibnamefont {Chow}}, \bibinfo
  {author} {\bibfnamefont {S.~T.}\ \bibnamefont {Merkel}}, \bibinfo {author}
  {\bibfnamefont {M.~P.}\ \bibnamefont {Da~Silva}}, \bibinfo {author}
  {\bibfnamefont {G.~A.}\ \bibnamefont {Keefe}}, \bibinfo {author}
  {\bibfnamefont {M.~B.}\ \bibnamefont {Rothwell}}, \bibinfo {author}
  {\bibfnamefont {T.~A.}\ \bibnamefont {Ohki}}, \bibinfo {author}
  {\bibfnamefont {M.~B.}\ \bibnamefont {Ketchen}},\ and\ \bibinfo {author}
  {\bibfnamefont {M.}~\bibnamefont {Steffen}},\ }\bibfield  {title} {\bibinfo
  {title} {{Efficient measurement of quantum gate error by interleaved
  randomized benchmarking}},\ }\bibfield  {journal} {\bibinfo  {journal}
  {Physical Review Letters}\ }\textbf {\bibinfo {volume} {109}},\ \href
  {https://doi.org/10.1103/PhysRevLett.109.080505}
  {10.1103/PhysRevLett.109.080505} (\bibinfo {year} {2012})\BibitemShut
  {NoStop}%
\bibitem [{\citenamefont {Bermudez}\ \emph {et~al.}(2017)\citenamefont
  {Bermudez}, \citenamefont {Xu}, \citenamefont {Nigmatullin}, \citenamefont
  {O’Gorman}, \citenamefont {Negnevitsky}, \citenamefont {Schindler},
  \citenamefont {Monz}, \citenamefont {Poschinger}, \citenamefont {Hempel},
  \citenamefont {Home}, \citenamefont {Schmidt-Kaler}, \citenamefont {Biercuk},
  \citenamefont {Blatt}, \citenamefont {Benjamin},\ and\ \citenamefont
  {M{\"{u}}ller}}]{Bermudez2017AssessingComputation}%
  \BibitemOpen
  \bibfield  {author} {\bibinfo {author} {\bibfnamefont {A.}~\bibnamefont
  {Bermudez}}, \bibinfo {author} {\bibfnamefont {X.}~\bibnamefont {Xu}},
  \bibinfo {author} {\bibfnamefont {R.}~\bibnamefont {Nigmatullin}}, \bibinfo
  {author} {\bibfnamefont {J.}~\bibnamefont {O’Gorman}}, \bibinfo {author}
  {\bibfnamefont {V.}~\bibnamefont {Negnevitsky}}, \bibinfo {author}
  {\bibfnamefont {P.}~\bibnamefont {Schindler}}, \bibinfo {author}
  {\bibfnamefont {T.}~\bibnamefont {Monz}}, \bibinfo {author} {\bibfnamefont
  {U.}~\bibnamefont {Poschinger}}, \bibinfo {author} {\bibfnamefont
  {C.}~\bibnamefont {Hempel}}, \bibinfo {author} {\bibfnamefont
  {J.}~\bibnamefont {Home}}, \bibinfo {author} {\bibfnamefont {F.}~\bibnamefont
  {Schmidt-Kaler}}, \bibinfo {author} {\bibfnamefont {M.}~\bibnamefont
  {Biercuk}}, \bibinfo {author} {\bibfnamefont {R.}~\bibnamefont {Blatt}},
  \bibinfo {author} {\bibfnamefont {S.}~\bibnamefont {Benjamin}},\ and\
  \bibinfo {author} {\bibfnamefont {M.}~\bibnamefont {M{\"{u}}ller}},\
  }\bibfield  {title} {\bibinfo {title} {{Assessing the Progress of Trapped-Ion
  Processors Towards Fault-Tolerant Quantum Computation}},\ }\href
  {https://doi.org/10.1103/PhysRevX.7.041061} {\bibfield  {journal} {\bibinfo
  {journal} {Physical Review X}\ }\textbf {\bibinfo {volume} {7}},\ \bibinfo
  {pages} {041061} (\bibinfo {year} {2017})}\BibitemShut {NoStop}%
\bibitem [{\citenamefont {Feng}\ \emph {et~al.}(2016)\citenamefont {Feng},
  \citenamefont {Wallman}, \citenamefont {Buonacorsi}, \citenamefont {Cho},
  \citenamefont {Park}, \citenamefont {Xin}, \citenamefont {Lu}, \citenamefont
  {Baugh},\ and\ \citenamefont {Laflamme}}]{Feng2016EstimatingQubit}%
  \BibitemOpen
  \bibfield  {author} {\bibinfo {author} {\bibfnamefont {G.}~\bibnamefont
  {Feng}}, \bibinfo {author} {\bibfnamefont {J.~J.}\ \bibnamefont {Wallman}},
  \bibinfo {author} {\bibfnamefont {B.}~\bibnamefont {Buonacorsi}}, \bibinfo
  {author} {\bibfnamefont {F.~H.}\ \bibnamefont {Cho}}, \bibinfo {author}
  {\bibfnamefont {D.~K.}\ \bibnamefont {Park}}, \bibinfo {author}
  {\bibfnamefont {T.}~\bibnamefont {Xin}}, \bibinfo {author} {\bibfnamefont
  {D.}~\bibnamefont {Lu}}, \bibinfo {author} {\bibfnamefont {J.}~\bibnamefont
  {Baugh}},\ and\ \bibinfo {author} {\bibfnamefont {R.}~\bibnamefont
  {Laflamme}},\ }\bibfield  {title} {\bibinfo {title} {{Estimating the
  Coherence of Noise in Quantum Control of a Solid-State Qubit}},\ }\bibfield
  {journal} {\bibinfo  {journal} {Physical Review Letters}\ }\textbf {\bibinfo
  {volume} {117}},\ \href {https://doi.org/10.1103/PhysRevLett.117.260501}
  {10.1103/PhysRevLett.117.260501} (\bibinfo {year} {2016})\BibitemShut
  {NoStop}%
\bibitem [{\citenamefont {Wallman}\ \emph {et~al.}(2015)\citenamefont
  {Wallman}, \citenamefont {Granade}, \citenamefont {Harper},\ and\
  \citenamefont {Flammia}}]{Wallman2015EstimatingNoise}%
  \BibitemOpen
  \bibfield  {author} {\bibinfo {author} {\bibfnamefont {J.}~\bibnamefont
  {Wallman}}, \bibinfo {author} {\bibfnamefont {C.}~\bibnamefont {Granade}},
  \bibinfo {author} {\bibfnamefont {R.}~\bibnamefont {Harper}},\ and\ \bibinfo
  {author} {\bibfnamefont {S.~T.}\ \bibnamefont {Flammia}},\ }\bibfield
  {title} {\bibinfo {title} {{Estimating the coherence of noise}},\ }\bibfield
  {journal} {\bibinfo  {journal} {New Journal of Physics}\ }\textbf {\bibinfo
  {volume} {17}},\ \href {https://doi.org/10.1088/1367-2630/17/11/113020}
  {10.1088/1367-2630/17/11/113020} (\bibinfo {year} {2015})\BibitemShut
  {NoStop}%
\bibitem [{\citenamefont {Chen}(2018)}]{Chen2018MetrologyQubits}%
  \BibitemOpen
  \bibfield  {author} {\bibinfo {author} {\bibfnamefont {Z.}~\bibnamefont
  {Chen}},\ }\emph {\bibinfo {title} {Metrology of Quantum Control and
  Measurement in Superconducting Qubits}},\ \href@noop {} {Ph.D. thesis}
  (\bibinfo {year} {2018})\BibitemShut {NoStop}%
\bibitem [{\citenamefont {Yeh}\ \emph {et~al.}(2017)\citenamefont {Yeh},
  \citenamefont {Lefebvre}, \citenamefont {Premaratne}, \citenamefont
  {Wellstood},\ and\ \citenamefont {Palmer}}]{Yeh2017MicrowaveMK}%
  \BibitemOpen
  \bibfield  {author} {\bibinfo {author} {\bibfnamefont {J.~H.}\ \bibnamefont
  {Yeh}}, \bibinfo {author} {\bibfnamefont {J.}~\bibnamefont {Lefebvre}},
  \bibinfo {author} {\bibfnamefont {S.}~\bibnamefont {Premaratne}}, \bibinfo
  {author} {\bibfnamefont {F.~C.}\ \bibnamefont {Wellstood}},\ and\ \bibinfo
  {author} {\bibfnamefont {B.~S.}\ \bibnamefont {Palmer}},\ }\bibfield  {title}
  {\bibinfo {title} {{Microwave attenuators for use with quantum devices below
  100 mK}},\ }\bibfield  {journal} {\bibinfo  {journal} {Journal of Applied
  Physics}\ }\textbf {\bibinfo {volume} {121}},\ \href
  {https://doi.org/10.1063/1.4984894} {10.1063/1.4984894} (\bibinfo {year}
  {2017})\BibitemShut {NoStop}%
\bibitem [{\citenamefont {Wang}\ \emph {et~al.}(2019)\citenamefont {Wang},
  \citenamefont {Shankar}, \citenamefont {Minev}, \citenamefont
  {Campagne-Ibarcq}, \citenamefont {Narla},\ and\ \citenamefont
  {Devoret}}]{Wang2019CavityQubits}%
  \BibitemOpen
  \bibfield  {author} {\bibinfo {author} {\bibfnamefont {Z.}~\bibnamefont
  {Wang}}, \bibinfo {author} {\bibfnamefont {S.}~\bibnamefont {Shankar}},
  \bibinfo {author} {\bibfnamefont {Z.~K.}\ \bibnamefont {Minev}}, \bibinfo
  {author} {\bibfnamefont {P.}~\bibnamefont {Campagne-Ibarcq}}, \bibinfo
  {author} {\bibfnamefont {A.}~\bibnamefont {Narla}},\ and\ \bibinfo {author}
  {\bibfnamefont {M.~H.}\ \bibnamefont {Devoret}},\ }\bibfield  {title}
  {\bibinfo {title} {{Cavity Attenuators for Superconducting Qubits}},\
  }\bibfield  {journal} {\bibinfo  {journal} {Physical Review Applied}\
  }\textbf {\bibinfo {volume} {11}},\ \href
  {https://doi.org/10.1103/PhysRevApplied.11.014031}
  {10.1103/PhysRevApplied.11.014031} (\bibinfo {year} {2019})\BibitemShut
  {NoStop}%
\bibitem [{\citenamefont {Schuster}\ \emph {et~al.}(2005)\citenamefont
  {Schuster}, \citenamefont {Wallraff}, \citenamefont {Blais}, \citenamefont
  {Frunzio}, \citenamefont {Huang}, \citenamefont {Majer}, \citenamefont
  {Girvin},\ and\ \citenamefont {Schoelkopf}}]{Schuster2005AcField}%
  \BibitemOpen
  \bibfield  {author} {\bibinfo {author} {\bibfnamefont {D.~I.}\ \bibnamefont
  {Schuster}}, \bibinfo {author} {\bibfnamefont {A.}~\bibnamefont {Wallraff}},
  \bibinfo {author} {\bibfnamefont {A.}~\bibnamefont {Blais}}, \bibinfo
  {author} {\bibfnamefont {L.}~\bibnamefont {Frunzio}}, \bibinfo {author}
  {\bibfnamefont {R.~S.}\ \bibnamefont {Huang}}, \bibinfo {author}
  {\bibfnamefont {J.}~\bibnamefont {Majer}}, \bibinfo {author} {\bibfnamefont
  {S.~M.}\ \bibnamefont {Girvin}},\ and\ \bibinfo {author} {\bibfnamefont
  {R.~J.}\ \bibnamefont {Schoelkopf}},\ }\bibfield  {title} {\bibinfo {title}
  {{Ac Stark shift and dephasing of a superconducting qubit strongly coupled to
  a cavity field}},\ }\bibfield  {journal} {\bibinfo  {journal} {Physical
  Review Letters}\ }\textbf {\bibinfo {volume} {94}},\ \href
  {https://doi.org/10.1103/PhysRevLett.94.123602}
  {10.1103/PhysRevLett.94.123602} (\bibinfo {year} {2005})\BibitemShut
  {NoStop}%
\bibitem [{\citenamefont {Jin}\ \emph {et~al.}(2015)\citenamefont {Jin},
  \citenamefont {Kamal}, \citenamefont {Sears}, \citenamefont {Gudmundsen},
  \citenamefont {Hover}, \citenamefont {Miloshi}, \citenamefont {Slattery},
  \citenamefont {Yan}, \citenamefont {Yoder}, \citenamefont {Orlando},
  \citenamefont {Gustavsson},\ and\ \citenamefont
  {Oliver}}]{Jin2015ThermalQubit}%
  \BibitemOpen
  \bibfield  {author} {\bibinfo {author} {\bibfnamefont {X.~Y.}\ \bibnamefont
  {Jin}}, \bibinfo {author} {\bibfnamefont {A.}~\bibnamefont {Kamal}}, \bibinfo
  {author} {\bibfnamefont {A.~P.}\ \bibnamefont {Sears}}, \bibinfo {author}
  {\bibfnamefont {T.}~\bibnamefont {Gudmundsen}}, \bibinfo {author}
  {\bibfnamefont {D.}~\bibnamefont {Hover}}, \bibinfo {author} {\bibfnamefont
  {J.}~\bibnamefont {Miloshi}}, \bibinfo {author} {\bibfnamefont
  {R.}~\bibnamefont {Slattery}}, \bibinfo {author} {\bibfnamefont
  {F.}~\bibnamefont {Yan}}, \bibinfo {author} {\bibfnamefont {J.}~\bibnamefont
  {Yoder}}, \bibinfo {author} {\bibfnamefont {T.~P.}\ \bibnamefont {Orlando}},
  \bibinfo {author} {\bibfnamefont {S.}~\bibnamefont {Gustavsson}},\ and\
  \bibinfo {author} {\bibfnamefont {W.~D.}\ \bibnamefont {Oliver}},\ }\bibfield
   {title} {\bibinfo {title} {{Thermal and Residual Excited-State Population in
  a 3D Transmon Qubit}},\ }\bibfield  {journal} {\bibinfo  {journal} {Physical
  Review Letters}\ }\textbf {\bibinfo {volume} {114}},\ \href
  {https://doi.org/10.1103/PhysRevLett.114.240501}
  {10.1103/PhysRevLett.114.240501} (\bibinfo {year} {2015})\BibitemShut
  {NoStop}%
\bibitem [{\citenamefont {Pop}\ \emph {et~al.}(2014)\citenamefont {Pop},
  \citenamefont {Geerlings}, \citenamefont {Catelani}, \citenamefont
  {Schoelkopf}, \citenamefont {Glazman},\ and\ \citenamefont
  {Devoret}}]{Pop2014CoherentQuasiparticles}%
  \BibitemOpen
  \bibfield  {author} {\bibinfo {author} {\bibfnamefont {I.~M.}\ \bibnamefont
  {Pop}}, \bibinfo {author} {\bibfnamefont {K.}~\bibnamefont {Geerlings}},
  \bibinfo {author} {\bibfnamefont {G.}~\bibnamefont {Catelani}}, \bibinfo
  {author} {\bibfnamefont {R.~J.}\ \bibnamefont {Schoelkopf}}, \bibinfo
  {author} {\bibfnamefont {L.~I.}\ \bibnamefont {Glazman}},\ and\ \bibinfo
  {author} {\bibfnamefont {M.~H.}\ \bibnamefont {Devoret}},\ }\bibfield
  {title} {\bibinfo {title} {{Coherent suppression of electromagnetic
  dissipation due to superconducting quasiparticles}},\ }\href
  {https://doi.org/10.1038/nature13017} {\bibfield  {journal} {\bibinfo
  {journal} {Nature}\ }\textbf {\bibinfo {volume} {508}},\ \bibinfo {pages}
  {369} (\bibinfo {year} {2014})}\BibitemShut {NoStop}%
\bibitem [{\citenamefont {Klimov}\ \emph {et~al.}(2018)\citenamefont {Klimov},
  \citenamefont {Kelly}, \citenamefont {Chen}, \citenamefont {Neeley},
  \citenamefont {Megrant}, \citenamefont {Burkett}, \citenamefont {Barends},
  \citenamefont {Arya}, \citenamefont {Chiaro}, \citenamefont {Chen},
  \citenamefont {Dunsworth}, \citenamefont {Fowler}, \citenamefont {Foxen},
  \citenamefont {Gidney}, \citenamefont {Giustina}, \citenamefont {Graff},
  \citenamefont {Huang}, \citenamefont {Jeffrey}, \citenamefont {Lucero},
  \citenamefont {Mutus}, \citenamefont {Naaman}, \citenamefont {Neill},
  \citenamefont {Quintana}, \citenamefont {Roushan}, \citenamefont {Sank},
  \citenamefont {Vainsencher}, \citenamefont {Wenner}, \citenamefont {White},
  \citenamefont {Boixo}, \citenamefont {Babbush}, \citenamefont {Smelyanskiy},
  \citenamefont {Neven},\ and\ \citenamefont
  {Martinis}}]{Klimov2018FluctuationsQubits}%
  \BibitemOpen
  \bibfield  {author} {\bibinfo {author} {\bibfnamefont {P.~V.}\ \bibnamefont
  {Klimov}}, \bibinfo {author} {\bibfnamefont {J.}~\bibnamefont {Kelly}},
  \bibinfo {author} {\bibfnamefont {Z.}~\bibnamefont {Chen}}, \bibinfo {author}
  {\bibfnamefont {M.}~\bibnamefont {Neeley}}, \bibinfo {author} {\bibfnamefont
  {A.}~\bibnamefont {Megrant}}, \bibinfo {author} {\bibfnamefont
  {B.}~\bibnamefont {Burkett}}, \bibinfo {author} {\bibfnamefont
  {R.}~\bibnamefont {Barends}}, \bibinfo {author} {\bibfnamefont
  {K.}~\bibnamefont {Arya}}, \bibinfo {author} {\bibfnamefont {B.}~\bibnamefont
  {Chiaro}}, \bibinfo {author} {\bibfnamefont {Y.}~\bibnamefont {Chen}},
  \bibinfo {author} {\bibfnamefont {A.}~\bibnamefont {Dunsworth}}, \bibinfo
  {author} {\bibfnamefont {A.}~\bibnamefont {Fowler}}, \bibinfo {author}
  {\bibfnamefont {B.}~\bibnamefont {Foxen}}, \bibinfo {author} {\bibfnamefont
  {C.}~\bibnamefont {Gidney}}, \bibinfo {author} {\bibfnamefont
  {M.}~\bibnamefont {Giustina}}, \bibinfo {author} {\bibfnamefont
  {R.}~\bibnamefont {Graff}}, \bibinfo {author} {\bibfnamefont
  {T.}~\bibnamefont {Huang}}, \bibinfo {author} {\bibfnamefont
  {E.}~\bibnamefont {Jeffrey}}, \bibinfo {author} {\bibfnamefont
  {E.}~\bibnamefont {Lucero}}, \bibinfo {author} {\bibfnamefont {J.~Y.}\
  \bibnamefont {Mutus}}, \bibinfo {author} {\bibfnamefont {O.}~\bibnamefont
  {Naaman}}, \bibinfo {author} {\bibfnamefont {C.}~\bibnamefont {Neill}},
  \bibinfo {author} {\bibfnamefont {C.}~\bibnamefont {Quintana}}, \bibinfo
  {author} {\bibfnamefont {P.}~\bibnamefont {Roushan}}, \bibinfo {author}
  {\bibfnamefont {D.}~\bibnamefont {Sank}}, \bibinfo {author} {\bibfnamefont
  {A.}~\bibnamefont {Vainsencher}}, \bibinfo {author} {\bibfnamefont
  {J.}~\bibnamefont {Wenner}}, \bibinfo {author} {\bibfnamefont {T.~C.}\
  \bibnamefont {White}}, \bibinfo {author} {\bibfnamefont {S.}~\bibnamefont
  {Boixo}}, \bibinfo {author} {\bibfnamefont {R.}~\bibnamefont {Babbush}},
  \bibinfo {author} {\bibfnamefont {V.~N.}\ \bibnamefont {Smelyanskiy}},
  \bibinfo {author} {\bibfnamefont {H.}~\bibnamefont {Neven}},\ and\ \bibinfo
  {author} {\bibfnamefont {J.~M.}\ \bibnamefont {Martinis}},\ }\bibfield
  {title} {\bibinfo {title} {{Fluctuations of Energy-Relaxation Times in
  Superconducting Qubits}},\ }\bibfield  {journal} {\bibinfo  {journal}
  {Physical Review Letters}\ }\textbf {\bibinfo {volume} {121}},\ \href
  {https://doi.org/10.1103/PhysRevLett.121.090502}
  {10.1103/PhysRevLett.121.090502} (\bibinfo {year} {2018})\BibitemShut
  {NoStop}%
\bibitem [{\citenamefont {Catelani}\ \emph {et~al.}(2011)\citenamefont
  {Catelani}, \citenamefont {Schoelkopf}, \citenamefont {Devoret},\ and\
  \citenamefont {Glazman}}]{Catelani2011RelaxationQubits}%
  \BibitemOpen
  \bibfield  {author} {\bibinfo {author} {\bibfnamefont {G.}~\bibnamefont
  {Catelani}}, \bibinfo {author} {\bibfnamefont {R.~J.}\ \bibnamefont
  {Schoelkopf}}, \bibinfo {author} {\bibfnamefont {M.~H.}\ \bibnamefont
  {Devoret}},\ and\ \bibinfo {author} {\bibfnamefont {L.~I.}\ \bibnamefont
  {Glazman}},\ }\bibfield  {title} {\bibinfo {title} {{Relaxation and frequency
  shifts induced by quasiparticles in superconducting qubits}},\ }\href
  {https://doi.org/10.1103/PhysRevB.84.064517} {\bibfield  {journal} {\bibinfo
  {journal} {Physical Review B}\ }\textbf {\bibinfo {volume} {84}},\ \bibinfo
  {pages} {064517} (\bibinfo {year} {2011})}\BibitemShut {NoStop}%
\bibitem [{\citenamefont {Glazman}\ and\ \citenamefont
  {Catelani}(2020)}]{Glazman2020BogoliubovQubits}%
  \BibitemOpen
  \bibfield  {author} {\bibinfo {author} {\bibfnamefont {L.~I.}\ \bibnamefont
  {Glazman}}\ and\ \bibinfo {author} {\bibfnamefont {G.}~\bibnamefont
  {Catelani}},\ }\bibfield  {title} {\bibinfo {title} {{Bogoliubov
  Quasiparticles in Superconducting Qubits}},\ }\href
  {https://arxiv.org/abs/2003.04366} {\bibfield  {journal} {\bibinfo  {journal}
  {arXiv:2003.04366}\ } (\bibinfo {year} {2020})}\BibitemShut {NoStop}%
\bibitem [{\citenamefont {Houzet}\ \emph {et~al.}(2019)\citenamefont {Houzet},
  \citenamefont {Serniak}, \citenamefont {Catelani}, \citenamefont {Devoret},\
  and\ \citenamefont {Glazman}}]{Houzet2019Photon-assistedQubit}%
  \BibitemOpen
  \bibfield  {author} {\bibinfo {author} {\bibfnamefont {M.}~\bibnamefont
  {Houzet}}, \bibinfo {author} {\bibfnamefont {K.}~\bibnamefont {Serniak}},
  \bibinfo {author} {\bibfnamefont {G.}~\bibnamefont {Catelani}}, \bibinfo
  {author} {\bibfnamefont {M.~H.}\ \bibnamefont {Devoret}},\ and\ \bibinfo
  {author} {\bibfnamefont {L.~I.}\ \bibnamefont {Glazman}},\ }\bibfield
  {title} {\bibinfo {title} {{Photon-assisted charge-parity jumps in a
  superconducting qubit}},\ }\bibfield  {journal} {\bibinfo  {journal}
  {Physical Review Letters}\ }\textbf {\bibinfo {volume} {123}},\ \href
  {https://doi.org/10.1103/PhysRevLett.123.107704}
  {10.1103/PhysRevLett.123.107704} (\bibinfo {year} {2019})\BibitemShut
  {NoStop}%
\bibitem [{\citenamefont {Martinis}(2020)}]{Martinis2020SavingRays}%
  \BibitemOpen
  \bibfield  {author} {\bibinfo {author} {\bibfnamefont {J.~M.}\ \bibnamefont
  {Martinis}},\ }\bibfield  {title} {\bibinfo {title} {{Saving superconducting
  quantum processors from qubit decay and correlated errors generated by gamma
  and cosmic rays}},\ }\href {https://arxiv.org/abs/2012.06137} {\bibfield
  {journal} {\bibinfo  {journal} {ArXiv:2012.06137}\ } (\bibinfo {year}
  {2020})}\BibitemShut {NoStop}%
\bibitem [{\citenamefont {Veps{\"{a}}l{\"{a}}inen}\ \emph
  {et~al.}(2020)\citenamefont {Veps{\"{a}}l{\"{a}}inen}, \citenamefont
  {Karamlou}, \citenamefont {Orrell}, \citenamefont {Dogra}, \citenamefont
  {Loer}, \citenamefont {Vasconcelos}, \citenamefont {Kim}, \citenamefont
  {Melville}, \citenamefont {Niedzielski}, \citenamefont {Yoder}, \citenamefont
  {Gustavsson}, \citenamefont {Formaggio}, \citenamefont {VanDevender},\ and\
  \citenamefont {Oliver}}]{Vepsalainen2020ImpactCoherence}%
  \BibitemOpen
  \bibfield  {author} {\bibinfo {author} {\bibfnamefont {A.~P.}\ \bibnamefont
  {Veps{\"{a}}l{\"{a}}inen}}, \bibinfo {author} {\bibfnamefont {A.~H.}\
  \bibnamefont {Karamlou}}, \bibinfo {author} {\bibfnamefont {J.~L.}\
  \bibnamefont {Orrell}}, \bibinfo {author} {\bibfnamefont {A.~S.}\
  \bibnamefont {Dogra}}, \bibinfo {author} {\bibfnamefont {B.}~\bibnamefont
  {Loer}}, \bibinfo {author} {\bibfnamefont {F.}~\bibnamefont {Vasconcelos}},
  \bibinfo {author} {\bibfnamefont {D.~K.}\ \bibnamefont {Kim}}, \bibinfo
  {author} {\bibfnamefont {A.~J.}\ \bibnamefont {Melville}}, \bibinfo {author}
  {\bibfnamefont {B.~M.}\ \bibnamefont {Niedzielski}}, \bibinfo {author}
  {\bibfnamefont {J.~L.}\ \bibnamefont {Yoder}}, \bibinfo {author}
  {\bibfnamefont {S.}~\bibnamefont {Gustavsson}}, \bibinfo {author}
  {\bibfnamefont {J.~A.}\ \bibnamefont {Formaggio}}, \bibinfo {author}
  {\bibfnamefont {B.~A.}\ \bibnamefont {VanDevender}},\ and\ \bibinfo {author}
  {\bibfnamefont {W.~D.}\ \bibnamefont {Oliver}},\ }\bibfield  {title}
  {\bibinfo {title} {{Impact of ionizing radiation on superconducting qubit
  coherence}},\ }\href {https://doi.org/10.1038/s41586-020-2619-8} {\bibfield
  {journal} {\bibinfo  {journal} {Nature}\ }\textbf {\bibinfo {volume} {584}},\
  \bibinfo {pages} {551} (\bibinfo {year} {2020})}\BibitemShut {NoStop}%
\end{thebibliography}

%

\end{document}


\title{Supplemental Materials for\\
`Millisecond coherence in a superconducting qubit'}

\author{A. Somoroff et al.}

\date{\today}

\maketitle

\null\newpage
\clearpage

\section{Supplementary note 1: Device Parameters}
Fluxonium obeys the Hamiltonian \cite{Manucharyan2009Fluxonium:Offsets,Nguyen2019High-CoherenceQubit,NguyenTowardProcessor}:
\begin{equation}
    \hat{H} = 4E_C \hat{n}^2 + \frac{1}{2} E_L\hat{\varphi}^2 - E_J\cos\left(\hat{\varphi} - 2\pi \frac{\Phi_e}{\Phi_0} \right)
    \label{eq;fluxoniumHamiltonian}
\end{equation} 

\noindent where $E_C / h = 1.08 \mathrm{~GHz}$, $E_L / h = 0.64 \mathrm{~GHz}$, and $E_J / h = 5.57 \mathrm{~GHz}$ are the charging energy, the inductive energy and the Josephson energy, respectively. These parameters are extracted by fitting the two-tone spectroscopy of the device to the numerical diagonalization of Eq.~(\ref{eq;fluxoniumHamiltonian}) (see Supplementary Figure~\ref{fig:FigA111}). Fit error for all three values is below $0.001 \mathrm{~GHz}$.

\begin{figure}[h]
	\centering
	\includegraphics[width=1.05\linewidth]{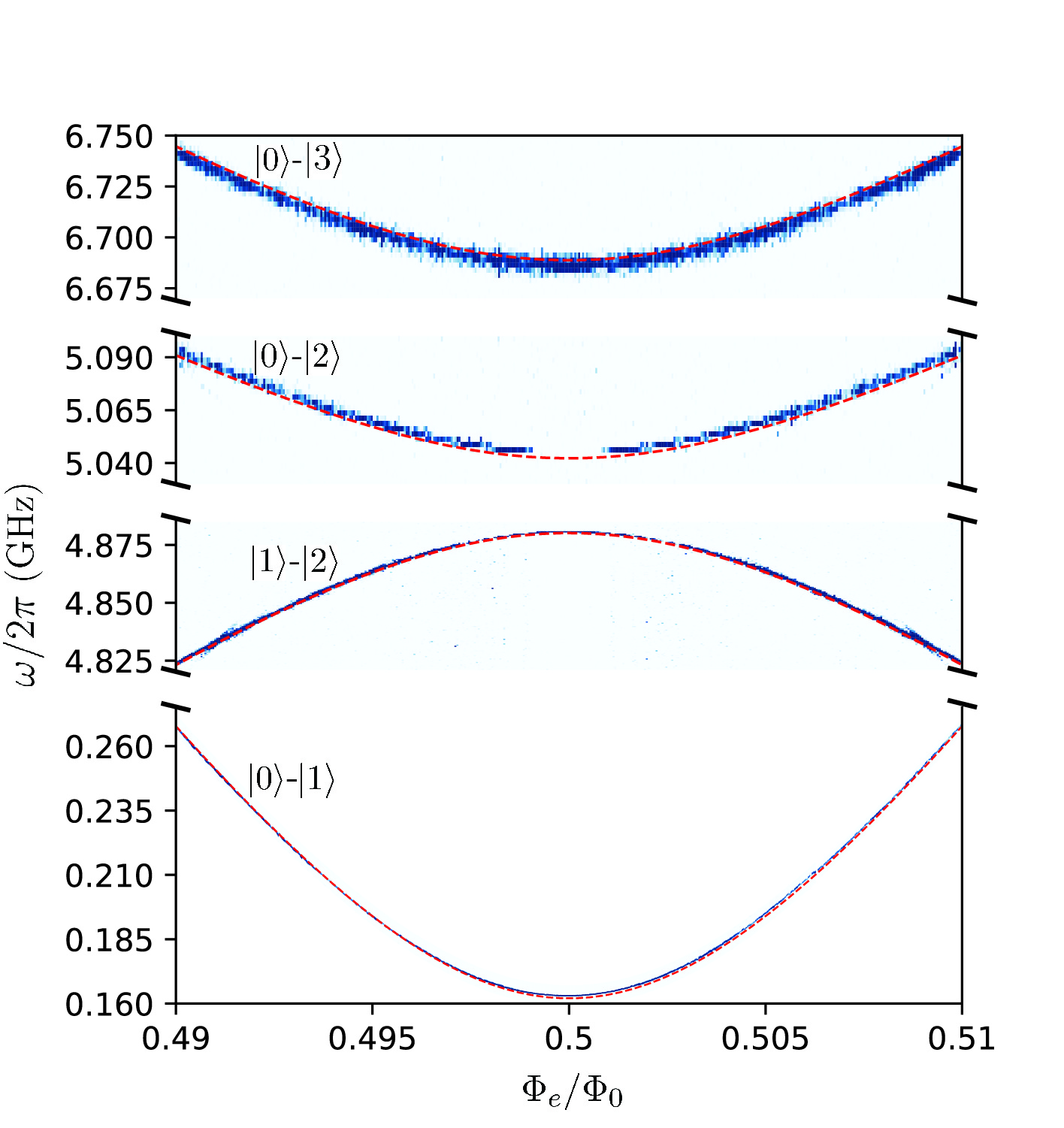}
	\caption{
	Two-tone spectroscopy of the fluxonium circuit as a function of the flux threading the loop (see Fig. 1a). The red dashed lines are the fits using Eq.~(\ref{eq;fluxoniumHamiltonian}).} 
	\label{fig:FigA111}
\end{figure}

\section{Supplementary note 2: Readout}

\subsection{Readout mode characterization}
\label{ro_characterize}
This device is embedded in a 3D copper cavity with a fundamental mode resonating at $7.54 \mathrm{~GHz}$. The qubit is dispersively coupled to the cavity mode with a coupling strength of $g/2\pi = 40$ MHz. From standard transmission measurements, we find the linewidth of the cavity mode $\kappa / 2\pi = 20.2$ MHz.

\begin{figure}[h]
	\centering
	\includegraphics[width=1.05\linewidth]{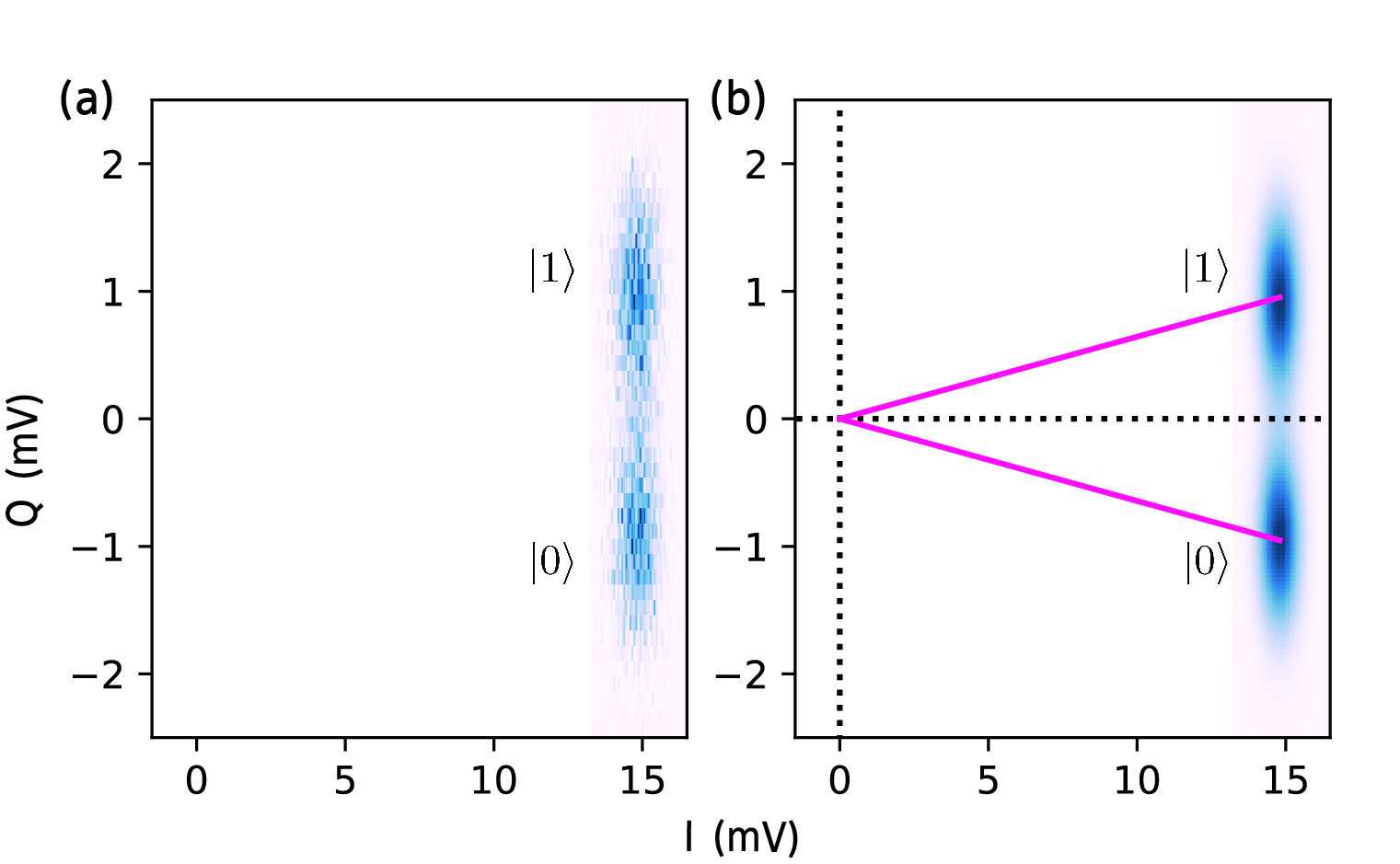}
	\caption{
	Experimental (a) and fitted (b) single shot histograms of the readout signal in the IQ plane. The histograms are calculated from 15,000 experimental realizations when the qubit starts in the thermal state. The angle between the magenta lines is used as a calibration of the ratio $\chi_{01}/\kappa=0.064$. This distribution is fitted to the sum of two Gaussians $A_0 e^{\frac{(x-x_0)^2}{\sigma^2}}+A_1 e^{\frac{(x-x_1)^2}{\sigma^2}}$. The relative height of the two gaussians encodes the population $p_0$ and $p_1$ of each qubit state according to $p_{0,1} = A_{0,1}/(A_1+A_0)$, giving the qubit temperature of $T= 25$ mK.}
	\label{fig:FigA1_1}
\end{figure}

We were able to perform a projective readout of the qubit state in a single-shot. Supplementary Figure \ref{fig:FigA1_1} shows the readout signal distribution in the complex plane. When the qubit is in the thermal state, we observe two separated blobs corresponding to the qubit being either in the $\left| 0 \right>$ or $\left| 1 \right>$ state. The angle between the two Blobs (magenta lines in Supplementary Figure~\ref{fig:FigA1_1}b) is used to calibrate $\chi_{01}/\kappa = 0.064$ and thus to extract the dispersive shift induced by the qubit on the cavity $\chi_{01}/2\pi = 1.3 \mathrm{~MHz}$.

\subsection{Thermal Cavity Photons}
\label{cav_photon}

The qubit dephasing rate due to thermal photons in the cavity is given by
\begin{equation}
    \Gamma_{\phi}^{\textup{th}} = \frac{n_{\textup{th}}\kappa \chi_{01}^2}{\kappa^2+\chi_{01}^2}
\end{equation}
in the low photon number $n_{\textup{th}}$ limit \cite{Zhang2017SuppressionQubit,Wang2019CavityQubits}. Plugging in the values of $\kappa$ and $\chi_{01}$ (see Supplementary Note \ref{ro_characterize}), a thermal dephasing time $T_{\phi}^{\textup{th}}=4.5$ ms corresponds to $n_{\textup{th}} \approx 0.0004$. From this value of $n_{\textup{th}}$ we determine that a cavity temperature of 45 mK is sufficient to explain the pure dephasing of our qubit.

\subsection{Readout Calibration}

\label{RO_Cal}

We observed that the readout pulse causes transitions between qubit states. To take into account this spurious effect, we calibrate the transition rate induced by applying a tone at the cavity frequency. In Supplementary Figure \ref{fig:FigA2}, we apply two successive cavity pulses. The first pulse is used to induce the transition between qubit state and reproduce what is happening during the readout with an increasing duration. The second pulse is used to readout qubit populations. 

\begin{figure}[h]
	\centering
	\includegraphics[width=\linewidth]{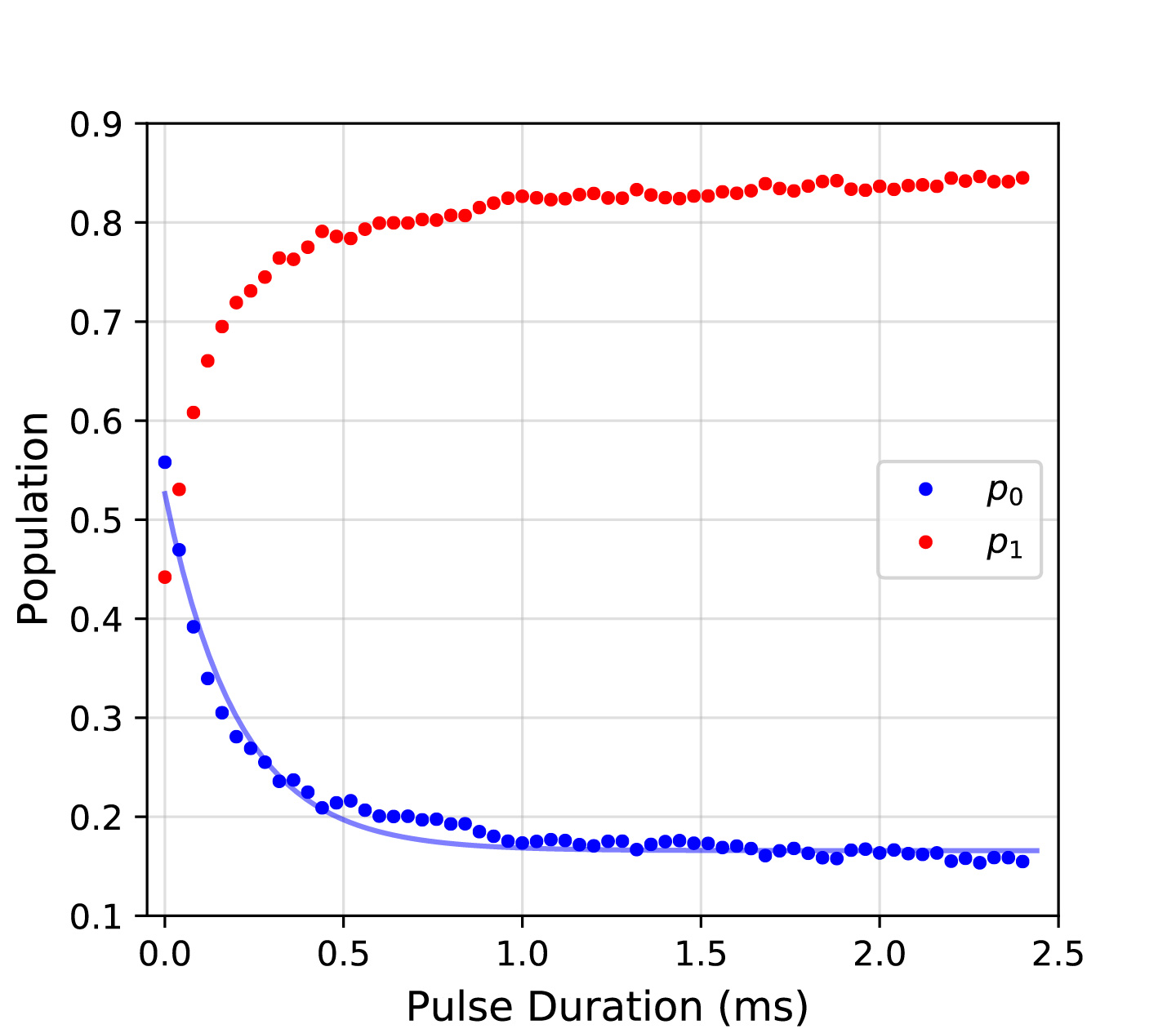}
	\caption{Qubit populations ($p_0$, $p_1$) after the application of a cavity pulse of varying duration. The populations are extracted from the single-shot measurement. The solid line is an exponential fit to $p_0(t)$ with a decay time of $0.204 \mathrm{~ms}$.}
	\label{fig:FigA2}
\end{figure} 

For simplicity, we will restrict our analysis to the ground state probability $p_0$.
In the presence of cavity photons, $p_0$ reads
\begin{equation}
    p_0(t) = [p_0(0) - p_0(\infty)]e^{-\frac{t}{T_1^{RO}}} + p_0(\infty)
\end{equation}
where $p_0(0),~ p_0(\infty)$ are $p_0$ at time equalling 0 and infinity, respectively, and $T_1^{RO}$ is the qubit relaxation time in the presence of cavity photons. The signal that we actually measure, $p_0^m$, is the average population during the duration of the readout pulse $T_{RO} = 20  ~ \mu \mathrm{s}$, given by:
\begin{equation}
    p_0^m = \frac{1}{T_{RO}}\int_{0}^{T_{RO}}p_0(t) dt.
\end{equation}

Integrating and solving for $p_0(0)$, we arrive at:
\begin{equation}
    p_0(0) = p_0(\infty) + \frac{p_0^m - p_0(\infty)}{\frac{T_1^{RO}}{T_{RO}}(1 - e^{-\frac{T_{RO}}{T_1^{RO}}})}.
\end{equation}
the parameters $p_0(\infty) = 0.166$, $T_1^{RO} = 204  ~ \mu \mathrm{s}$, and $p_0^m = 0.558$ are determined from the fit in Supplementary Figure \ref{fig:FigA2}.
With these values, we find that the thermal population in $p_0$, before any readout pulse is applied, is $p_0(0) = 0.578$, corresponding to an effective temperature of 25 mK. The operating temperature of the fridge was 8 mK.

In practice, we also use the transitions induced by the readout pulse to initialize the qubit into the $|1\rangle$ state with fidelity greater than 0.8, allowing for fast reset and improved contrast in measurements. In our experiments, such as the RB, we used a square cavity pulse of duration 160 $\mu\textup{s}$ and an amplitude 80\% of the normal readout pulse. 

\section{Supplementary note 3: Gate Tuning} 
\label{gate_tune}

The $X/2$ and $Y/2$ pulses were created using
Gaussian-edged square pulses with $60~\textrm{ns}$ plateau and $5~\textrm{ns}$-width edges totaling to an $80 \mathrm{~ns}$ duration. This pulse duration is chosen as it corresponds to a $\pi$ gate with nearly maximal power. No Derivative Removal by Adiabatic Gate (DRAG) correction or other state leakage mitigation technique was necessary thanks to the strong qubit anharmonicity. The relatively long plateau is an artifact of weak port coupling in our wireless setup in the $100~\textrm{MHz}$ frequency range. The $X$ ($Y$) pulses consist of pairs of concatenated $X/2$ ($Y/2)$ pulses, and they expectedly have about twice higher error (Fig.~3a, inset).

Fixing the time-domain parameters of the pulses, we tuned the amplitude of the pulses to find the optimal gate parameters. To do this, we performed a pulse train, where the qubit is driven with a sequence of $\pi/2$ pulses ranging from 0 to 120 pulses, with a step of 12. Each set of 12 $\pi/2$ pulses is equivalent to a $6\pi$ rotation of the qubit. For the correct pulse amplitude, the expected signal is flat while it should oscillate when the gate is imperfectly calibrated. We ran this pulse train, sweeping the pulse amplitude, and used the amplitude corresponding to the smallest standard deviation. This procedure was repeated four times, for plus and minus $\pi/2$ rotations around $X$ and $Y$ to take into account mixer imbalance and offset. We simply defined $\pi$ gates as two consecutive $\pi/2$ pulses with a 5 ns separation.

\section{Supplementary note 4: Dielectric Loss}
\label{diel}
\begin{figure}[h]
	\centering
	\includegraphics[width=\linewidth]{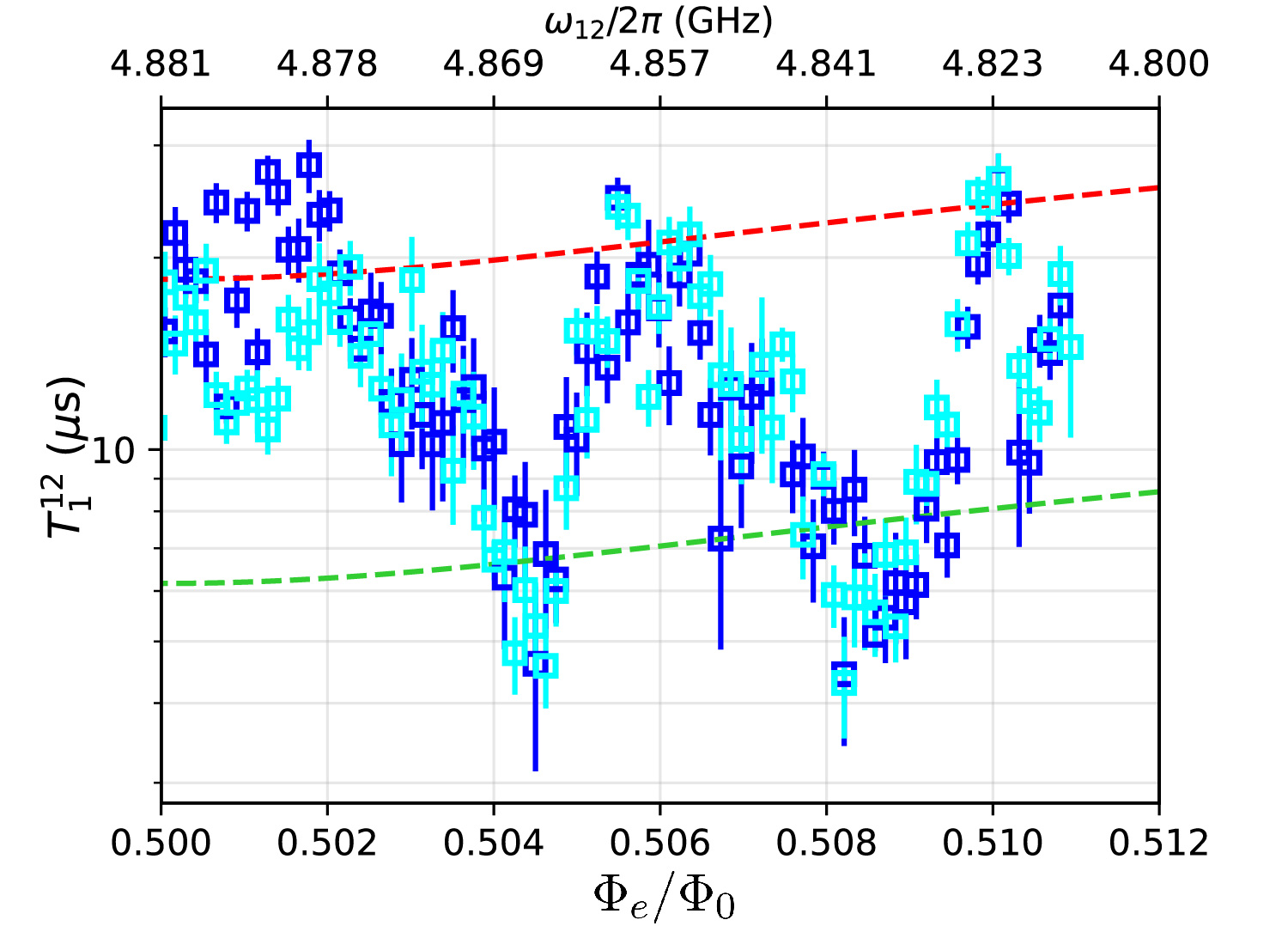}
	\caption{Energy relaxation time of the $|1\rangle-|2\rangle$ transition, $T_1^{12}$ as a function of the external magnetic flux. Dotted curves are $T_1$ limits imposed by dielectric loss for loss tangents of $1.5 \times 10^{-6}$ (red) and $4.5 \times 10^{-6}$ (green). Blue and light blue data points are the same scan taken twice.}
	\label{fig:FigA7}
\end{figure}

Material losses in the form of dielectric loss contribute to the qubit's energy relaxation time $T_1$ and therefore sets an upper limit on the maximal achievable coherence time $T_2$. In terms of the charge matrix element, $|\langle j|\hat{n}|i\rangle|$, the decay rate due to dielectic loss at a finite temperature T reads
\begin{equation}
    \Gamma_{ij}^{\mathrm{diel}} = 16 \pi (E_C/h)|\langle j|\hat{n}|i\rangle|^2 \tan \delta_C \left( 1 + \coth{\left(\frac{\hbar \omega_{ij}}{2k_B T} \right)} \right)
\end{equation}
where $\tan \delta_C$ is the dielectric loss tangent. We used this expression to calculate the upper bound imposed by dielectric losses in Fig.~4 and Supplementary Figure \ref{fig:FigA7}. 

\section{Supplementary note 5: Quasiparticles}
\label{Qp}
Quasiparticles tunneling across the the single Josephson junction and junctions within chain result in different decay rates on the qubit \cite{Serniak2018HotQubits,Pop2014CoherentQuasiparticles,Gustavsson2016SuppressingPumping}.

\subsection{Quasiparticle tunneling across the single Josephson junction}
The decay rate induced by quasiparticles tunneling across the small Josephson junction at zero temperature is
\begin{equation}
    \Gamma_{02}^{\mathrm{qp}} = \frac{16E_J}{\pi \hbar}\sqrt{\frac{2 \Delta}{\hbar \omega_{02}}}|\langle 2|\sin (\frac{\hat{\varphi} - \varphi_{e}}{2})|0\rangle|^2 x_{\mathrm{qp}} 
\end{equation}
where $\Delta \simeq 180 ~\mu \mathrm{eV}$ is the superconducting gap of Aluminum, $\varphi_e = 2\pi\frac{\Phi_e}{\Phi_0}$, and $x_{\mathrm{qp}}$ is the effective reduced quasiparticle density normalized by the number of cooper pairs.

\subsection{Quasiparticle tunneling across the chain of Josephson junctions}
The decay rate induced by quasiparticles tunneling across the Josephson junctions within the array providing the superinductance (shown in Fig.~1b) at zero temperature is

\begin{equation}
    \Gamma_{01}^{\mathrm{qp, array}} = \frac{16E_L}{\pi \hbar}\sqrt{\frac{2 \Delta}{\hbar \omega_{01}}}|\langle 1|\hat{\frac{\varphi}{2}}|0\rangle|^2 x_{\mathrm{qp}} 
\end{equation}
where $x_{\mathrm{qp}}$ is the quasiparticle density normalized by the number of cooper pairs. The value of $T_1^{01}$ at the sweet spot corresponds to $x_{\mathrm{qp}} = 6 \times 10^{-10}$. 

\section{Supplementary note 6: Relaxation and Coherence of the $|2\rangle - |1\rangle$ transition}

\label{t112}
To complete the study of the decay times of the qutrit comprised of fluxonium's three lowest energy levels, we also measured $T_1$ of the $|1\rangle - |2\rangle$ transition,  $T_1^{12}$. Like the $T_1^{01}$ data in figure 4, we observe "dips" in $T_1$ value, reproducible in time, which cannot be explained by tunneling of quasiparticles.

\section{Supplementary note 7: $T_1^{02}$ Measurement}
\label{t102}

\subsection{Experiment protocol}

In order to measure the relaxation rate associated with the $|2\rangle \rightarrow |0\rangle$ decay, we saturate the $|1\rangle-|2\rangle$ transition to even up the population $p_1$ and $p_2$, while monitoring the evolution of ground state population $p_0$ (see Supplementary Figure~\ref{fig:FigA4}a). The increase of population of the $\left| 0 \right>$ state is associated to relaxation events from state $\left| 2 \right>$ and state $\left| 1 \right>$. Owing to the large value of $T_1^{01}> 1 \mathrm{~ms}$, our measurement is extremely sensitive to low values of $T_1^{02} \ll T_1^{01}$. We record $p_0$ as a function of the duration of the $|1\rangle-|2\rangle$ drive and fit it to a decaying exponential. Supplementary Figure \ref{fig:FigA4}b shows example of traces obtained with our protocol. The saturation value as well as the decay constant enable us to determine $T_1^{02}$ unambiguously.

\begin{figure*}
	\centering
	\includegraphics[width=0.9\linewidth]{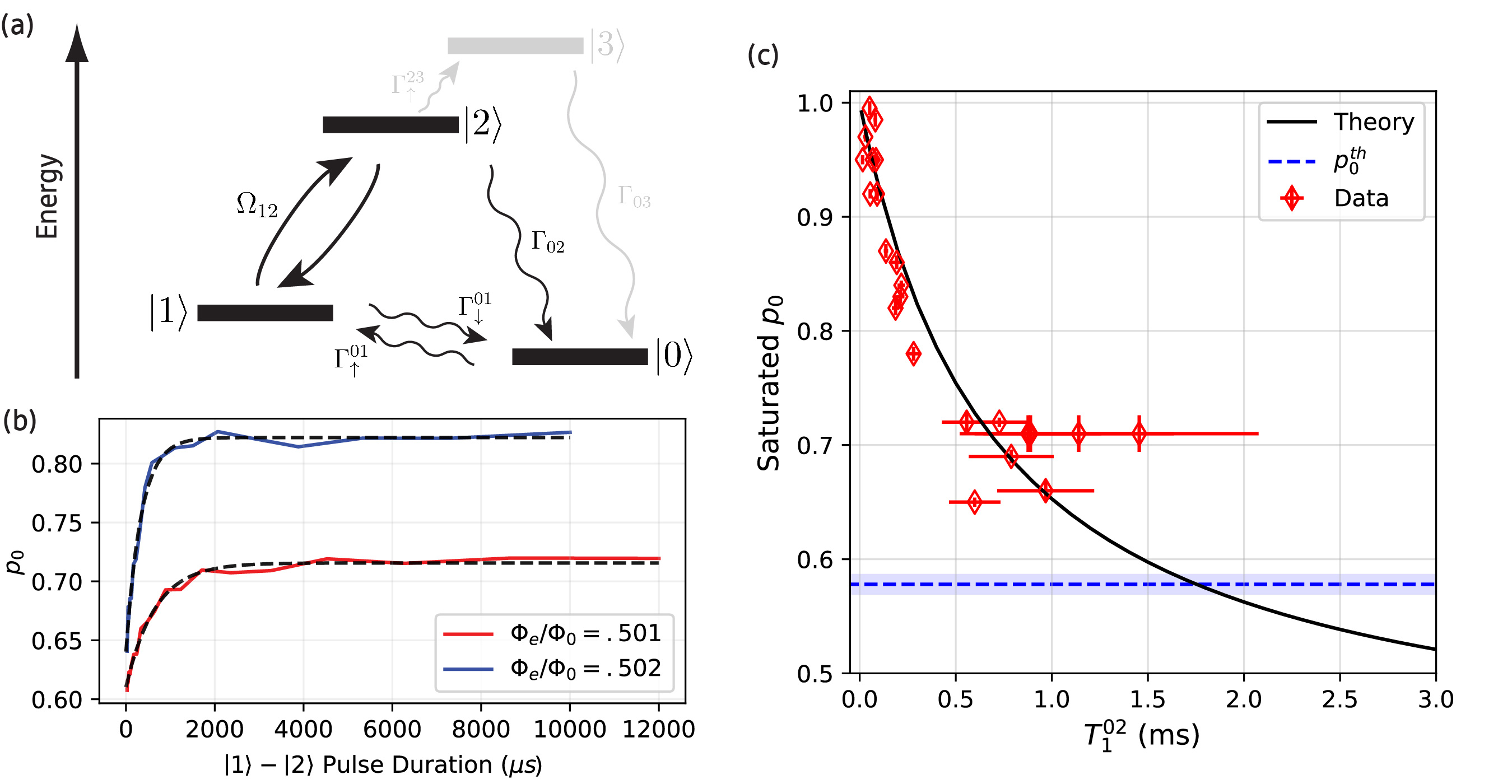}
	\caption{
	(a) Energy diagram of the first levels of the fluxonium. Wavy arrows represent incoherent transition rates while the curved arrows represent a coherent population transfer. We drive the $\left| 1 \right> - \left| 2 \right>$ with a Rabi frequency $\Omega_{12} \gg \Gamma_i$, faster than any relaxation rate of the system. We monitor the increase of $p_0$ caused by $|1\rangle \rightarrow |0\rangle$, $|2\rangle \rightarrow |0\rangle$, and $|2\rangle \rightarrow |3\rangle \rightarrow |0\rangle$ events. Transition events involving $|3\rangle$ are grey to signify that they are only relevant close to the sweet spot, where $T_1^{02} \sim 1 \mathrm{~ms}$ (see Supplementary Note \ref{23_excite}). (b) Experimental traces used to extract the lifetime $T_1^{02}$ in Fig.~4 b at two different flux points. For a lower $T_1^{02}$, $p_0$ saturates to a higher value. Blue and red curves are fitted to exponentials with time constants $(338 \pm 26)$ and $(651 \pm 48)$ $\mu$s, respectively. This time constant is used to extract $T_1^{02}$ according to equation (\ref{teff}). (c) Energy relaxation time $T_1^{02}$	with corresponding $p_0$ saturation value for each measurement of Fig.~4. The data agree well with the simulated curve using the theory derived in the previous section. When $T_1^{02}$ is in the range where the $p_0^{th}$ error bound intersects the black theory line, the precision is not sufficient to measure $T_1^{02}$. For $T_1^{02}$ above the intersection point, the protocol heats the qubit rather than cools it.}
	\label{fig:FigA4}
\end{figure*}

\subsection{Qutrit Model}

We outline our 3-level system (qutrit) model that we use to determine $T_1^{02}$ using the relaxation time of the experimental signal $T_{eff}$. Consider a qutrit in the presence of a continuous drive of the $|1\rangle - |2\rangle$ transition at frequency $\Omega_{12}$. We assume that $p_0 + p_1 + p_2 =1$, which is reasonable considering the temperature of the system and the transition frequencies. The dynamics of the three populations $p_0, p_1, p_2$ are given by the following linear system.

\begin{equation}
\begin{aligned}
&\frac{dp_0}{dt} = -\Gamma_{\uparrow}p_0 + \Gamma_{\downarrow}p_1 + \Gamma_{02}p_2 \\
&\frac{dp_1}{dt} = \Gamma_{\uparrow}p_0 - \Gamma_{\downarrow}p_1 + \Omega_{12}(p_2 - p_1)\\
&\frac{dp_2}{dt} = -\Gamma_{02}p_2 - \Omega_{12}(p_2 - p_1)
\end{aligned}
\end{equation}
\noindent
where $\Gamma_{\downarrow} = p_0^{th}\Gamma_{01}$ and $\Gamma_{\uparrow} = p_1^{th}\Gamma_{01}$ are the relaxation and excitation rates of the qubit transition. Note that $T_1^{01} = 1/\Gamma_{01} = 1/(\Gamma_{\downarrow} + \Gamma_{\uparrow})$. $p_0^{th}$, $p_1^{th}$ are the qubit populations at thermal equilibrium. 

Defining the variables $p_+ = p_1+p_2$ and $p_- = p_1-p_2$, we rewrite the system as
\begin{equation}
    \begin{aligned}
    & \frac{dp_0}{dt} = -\Gamma_{\uparrow}p_0 + \frac{\Gamma_{\downarrow}}{2}(p_-+p_+) - \frac{\Gamma_{02}}{2}(p_- - p_+)\\
    & \frac{dp_+}{dt} = \Gamma_{\uparrow}p_0 - \frac{\Gamma_{\downarrow}}{2}(p_-+p_+) + \frac{\Gamma_{02}}{2}(p_- - p_+)\\
    & \frac{dp_-}{dt} =  -\frac{\Gamma_{\downarrow}}{2}(p_-+p_+) + \Gamma_{\uparrow}p_0 - 2\Omega_{12}p_- +
    \frac{\Gamma_{02}}{2}(p_+ - p_-)
    \end{aligned}
\end{equation}

Because $\Omega_{12} \gg$ all $ \Gamma's$, we only keep the term $-2\Omega_{12}p_-$ in the last equation. Recall that the 12 drive $\Omega_{12}$ is applied for a duration long enough to even the populations in $\left| 1 \right>$ and $\left| 2 \right>$ so that $p_- \simeq 0$. We therefore obtain
\begin{equation}
    \begin{aligned}
    &\frac{dp_0}{dt} = -\Gamma_{\uparrow}p_0 + \frac{\Gamma_{\downarrow}+\Gamma_{02}}{2}p_+ \\
    &\frac{dp_+}{dt} = \Gamma_{\uparrow}p_0 - \frac{\Gamma_{\downarrow}+\Gamma_{02}}{2}p_+.
    \end{aligned}
\end{equation}

Inspecting this simplified linear system, we see that $p_0$ and $p_+$ exponentially saturate to their equilibrium values with the rate 
\begin{equation}
    \Gamma_{eff} = \Gamma_{\uparrow} + \frac{\Gamma_{\downarrow}+\Gamma_{02}}{2}.
\end{equation}

The effective decay constant that we extract from fitting $p_0$ as a function of $|1\rangle - |2\rangle$ drive duration in Supplementary Figure \ref{fig:FigA4} is therefore $ T_{eff} = 1/\Gamma_{eff}$. Solving for $\Gamma_{02}$, and converting the $\Gamma's$ to times, we arrive at the formula for $T_1^{02}$

\begin{equation}
\label{teff}
    T_1^{02} = \frac{T_1^{01}T_{eff}}{2T_1^{01} - (2 - p_0^{th})T_{eff}}.
\end{equation}

\subsection{$T_1^{02}$ at $\Phi_e/\Phi_0 = 0.5$}
\label{t1_02_hfq}

For relaxation times $T_1^{02} \ll T_1^{01}$, the protocol depicted above results in an increase of $p_0$. In this case, one can use it as a cooling scheme. We first saturate the $|1\rangle-|2\rangle$ transition and wait for population to accumulate in $\left| 0 \right>$. After turning off the drive and in a few $T_1^{12}$, the population is back to the computational subspace with a lower effective temperature. The cooling fidelity (i.e. the $p_0$ saturation value) depends on the ratio of $T_1^{01}$ to $T_1^{02}$. The smaller this ratio, the higher the cooling fidelity. For all of our measurements, $T_1^{01} \approx $ 1 ms. 
We have established through single shot measurements that $p_0^{th} = 0.580$ with a precision of 0.005. When the ratio
$T_1^{01}/T_1^{02} \sim 0.63$, the $p_0$ saturation value falls very close to $p_0^{th}$, such that the value will be within the error bound of $p_0^{th}$. In this case, we lack the precision to reliably measure $T_1^{02}.$ This phenomenon is outlined in Supplementary Figure \ref{fig:FigA4}c. Since $T_1^{02}$ given by Eq.~(\ref{teff}) seems to saturate to around 1.5 ms from the data points close to $\Phi_e/\Phi_0 = 0.5$ in figure 4, we estimate that that $T_1^{02} \gtrsim 1.6-1.8$ ms at the sweet spot, shown as the gray point in figure 4.

\subsection{Thermal Excitation of $|2\rangle \rightarrow |3\rangle$}
\label{23_excite}

Given the qubit temperature of 25 mK and the transition frequency  $f_{23} = 1.66 \mathrm{~GHz}$, thermal excitation events $2 \rightarrow 3$ are possible during the measurement (see Supplementary Figure~\ref{fig:FigA4}a). We estimate the relaxation rate $\Gamma_1^{23}$ assuming that it is limited by dielectric losses with loss tangent $1 \times 10^{-6}$ and the excitation probability with the Boltzmann factor to find $\Gamma_{\uparrow}^{23} = \frac{e^{-\hbar \omega_{23}/k_BT}}{1+e^{-\hbar \omega_{23}/k_BT}} \Gamma_1^{23} \sim (1 \mathrm{~ms})^{-1}$. We find that the saturation value obtained in Fig.~4 at $\Phi_e/\Phi_0 = 0.5$ is likely limited by this process while $T_1^{02}$ is limited by dielectric losses away from this flux point. The $\sim$ ms value reported here therefore serves as a lower bound on the energy relaxation time of the transition.

A similar analysis shows that the $|1\rangle - |2\rangle$ (4.88 GHz) and $|0\rangle - |3\rangle$ (6.69 GHz) transitions have a thermal excitation rate of $(250~ \mathrm{ms})^{-1}$, and $(7~ \mathrm{s})^{-1}$, respectively. We conclude that these thermal excitations have no effect on measured qubit coherence.

\section{Supplementary note 8: Fabrication Procedure}

Our device was fabricated on a 430 $\mu$m-thick sapphire substrate diced into a $9 \times 4$ mm chip. We follow the following fabrication recipe:

\begin{itemize}
\item \textbf{Cleaning:} The chip is prepared by sonicating it in acetone for 3 minutes, followed by another sonicating bath in isopropyl alcohol (IPA) for 3 minutes. Upon removal from the IPA sonicating bath, the chip is blown dry with $N_2$. 
\item \textbf{Resist Application:} 1 drop of MMA EL-13 electron-beam resist is applied to the chip, then spun at 5000 RPM for 1 minute. The resist is then baked on a hotplate for 1 minute at $180^{\circ}$C. Second layer of resist is applied: 1 drop of PMMA A3 electron beam resist spun at 4000 RPM for 1 minute before barking at $180^{\circ}$C for 30 minutes. 
\item \textbf{Anti-Charging Aluminum Deposition:} Because sapphire is an insulator, a thin conducting layer must be applied to the surface of the chip to avoid charge accumulation. 11 nm of Al is deposited on the chip at 1 nm/s using a Plassys deposition system.
\item \textbf{Electron Beam Lithography:} The circuit is written with 100 kV Elionix Electron Beam Lithography system, with a current of 1 nA. 
\item \textbf{Aluminum Etch:} The anti-charging layer must be removed before the mask is developed. The Al is etched in a 0.1 M potassium hydroxide (KOH) aqueous solution. Chip is left in the solution until the Al is visibly gone.
\item \textbf{Development:} Mask is developed for 2 minutes in a 3:1 IPA:DI solution at $6^{\circ}$C. Chip is lightly shaken back and forth by hand at around 1-2 Hz while in the developer.
\item \textbf{Deposition:} Chip is loaded into the Plassys deposition system and the loadlock is pumped on for 20 hours until the pressure reaches $1.3 \times 10^{-7}$ mBar before the deposition. The deposition is comprised of the following steps: 1. 20 second Ar etch at each deposition angle ($\pm 23.83^{\circ}$) 2. Deposit Ti into the chamber at .1 nm/s for 2 minutes 3. First Al deposition: 20 nm is deposited at 1 nm/s at an angle of $23.83^{\circ}$ 4. 10 minutes of oxidation at 100 mBar 5. Second Al deposition: 40 nm is deposited at 1 nm/s at an angle of $-23.83^{\circ}$ 6. 20 minutes of oxidation at 10 mBar (capping).
\item \textbf{Liftoff:} Chip is bathed in acetone for 3 hours at $60^{\circ}$C. Then it is sonicated in the acetone for 5 seconds, followed by 10 seconds of sonication in IPA. Finally, chip is blown dry with $N_2$.
\end{itemize}

\section{Supplementary note 9: Experimental Setup}

\subsection{Room Temperature Electronics}
\label{roomtemp}
\begin{figure}[H]
	\centering
	\includegraphics[width=\linewidth]{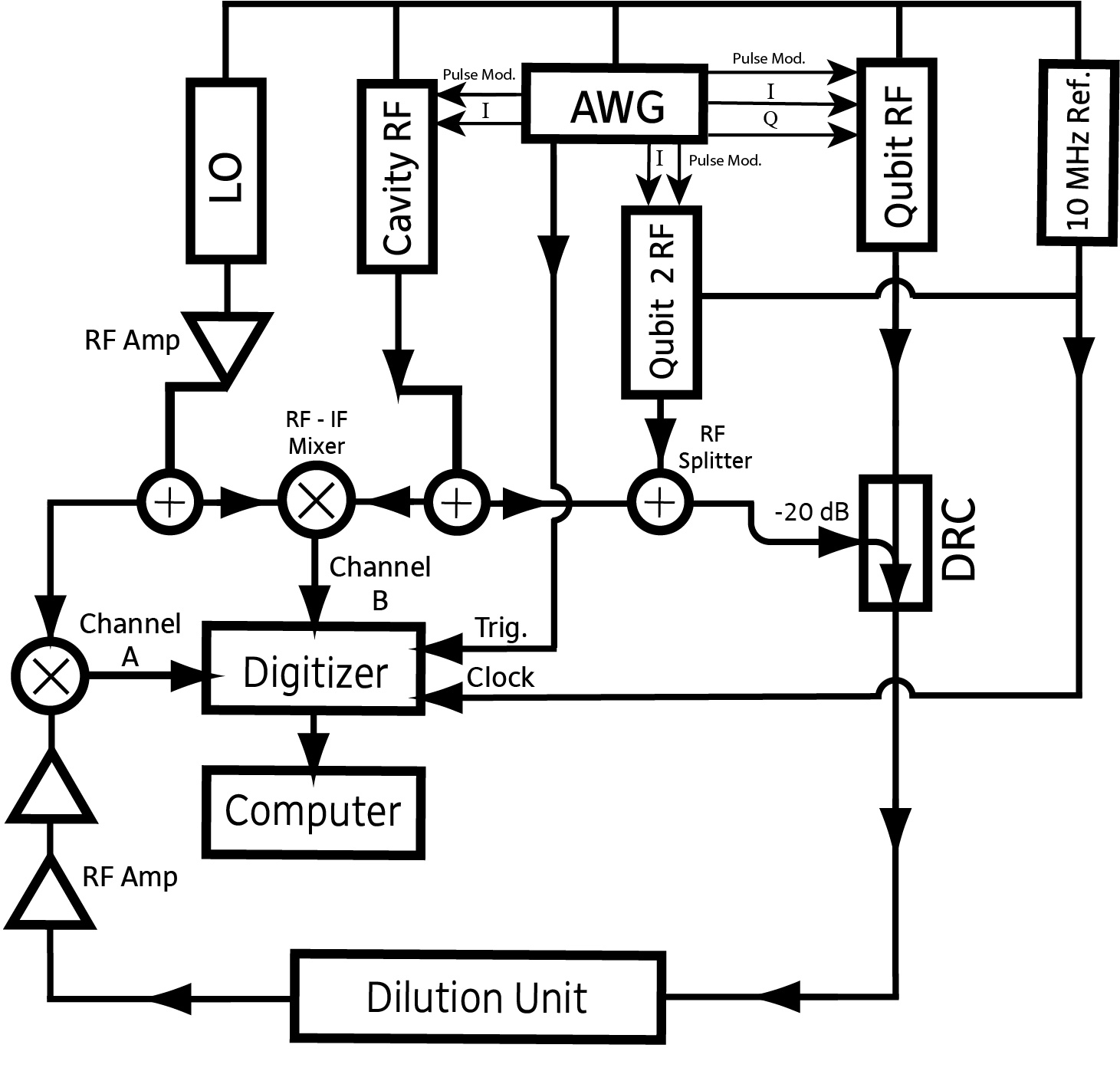}
	\caption{Schematic of the room temperature setup. Fluxonium $|0\rangle-|1\rangle$, $|1\rangle-|2\rangle$ transitions and cavity pulses are generated by a Tektronix Arbitrary Waveform Generator AWG5014C and up-converted using the mixing capabilities of three Rohde and Schwarz SMB100A generators dubbed "Qubit RF", "Qubit 2 RF", "Cavity RF". The signals are combined by a directional coupler to maximize the available qubit drive powers. The outgoing signal is amplified by room-temperature amplifiers before down-conversion by a HP8763B generator used as a local oscillator (LO), and digitization by an Alazar acquisition board.}
	\label{fig:FigA9}
\end{figure}

Supplementary Figure \ref{fig:FigA9} is a diagram of the room temperature setup. Pulses are provided and modulated with Rohde and Schwarz SMB100A generators and combined before entering the dilution unit. The signal from the output port of the refrigerator is amplified and mixed down with a local oscillator (LO) and finally digitized for numerical demodulation. A reference signal created by mixing the readout tone with the LO is acquired along with the signal. The phase of the two signals are subtracted to ensure the stability of the measurement setup.

\subsection{Cryogenic Setup}
\label{cryogenic}

Our experiments are performed in a dilution refrigerator with a base temperature $\sim 8 \mathrm{~mK}$. Our device is protected from thermal noise by a combination of attenuation and filters depicted in Supplementary Figure \ref{fig:FigA10}. The readout signal transmitted through the cavity is amplified using a HEMT. Two circulators and a low-pass filter are placed between the amplifier and the cavity to prevent the noise amplified by the HEMT from entering the cavity.

\begin{figure}[H]
	\centering
	\includegraphics[width=\linewidth]{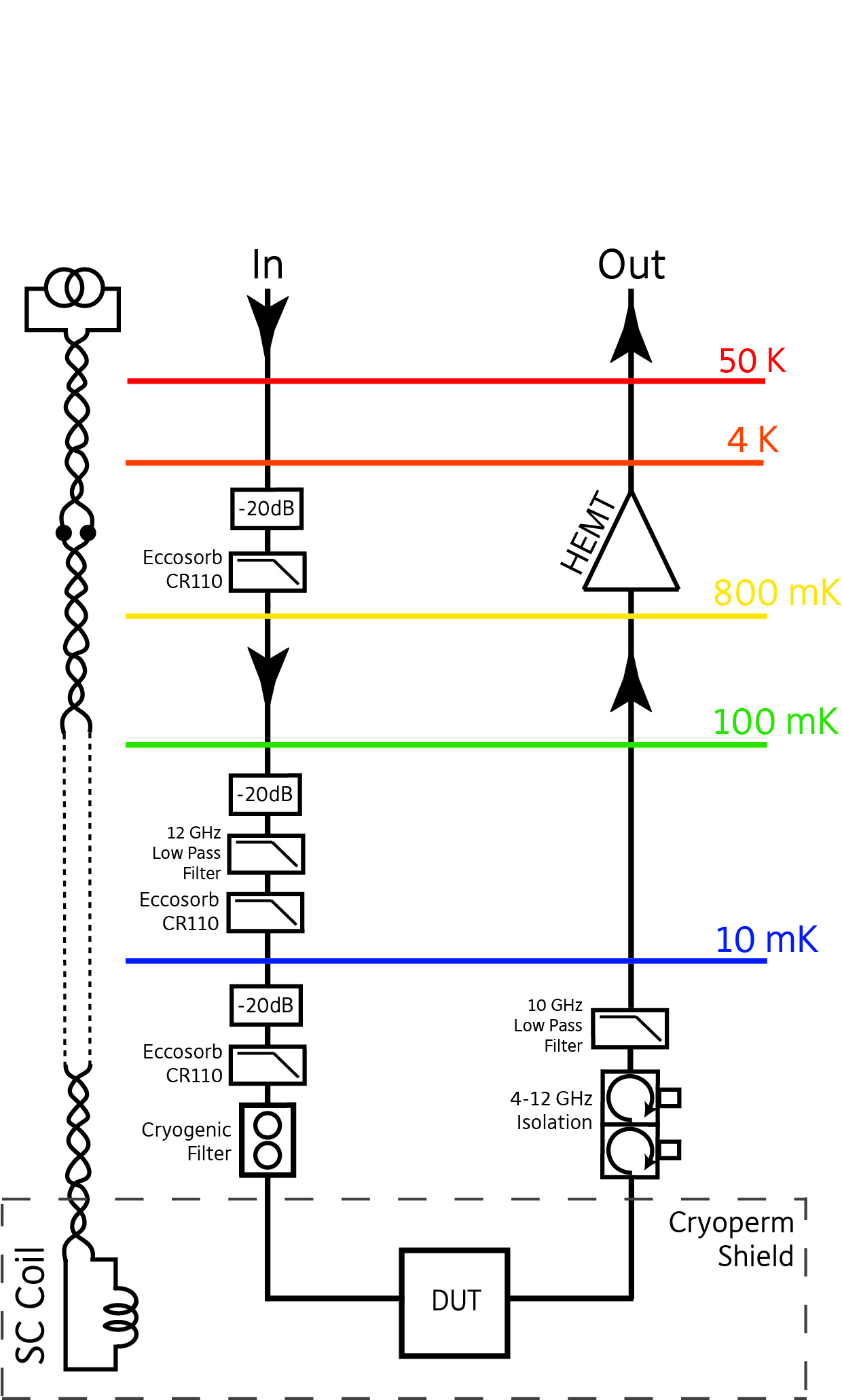}
	\caption{Schematic of the wiring of the fridge. The input line is heavily filtered using 60 dB of commercial attenuation along with home-made Eccosorb CR110 filters, and a 12 GHz K\&L low-pass filter. We add an home-made cryogenic low-pass filter as the last element before the input port of the cavity. The signal exiting the cavity is routed by two Low Noise Factory 4-12 GHz circulators, then a 10 GHz K\&L low-pass filter, before being amplified by a high electron mobility transistor (HEMT) at 4 K.}
	\label{fig:FigA10}
\end{figure} 	


%